\documentclass{aa}  

\usepackage[hyperindex=true,colorlinks=true,citecolor=blue,linkcolor=blue,breaklinks=true]{hyperref}
\usepackage{amsmath}
\usepackage{graphicx}
\usepackage{txfonts}
\usepackage{natbib,twoopt}
\usepackage{color}
\usepackage{multirow}
\usepackage{ulem}
\definecolor{red}{rgb}{1.,0.0,0.}

\definecolor{orange}{rgb}{1.,.65,0.}

\definecolor{vert}{rgb}{.0,.65,0.}

\newcommand{\kms}{$\mathrm{km\,s^{-1}}\,$}
\newcommand{\ms}{$\mathrm{m\,s^{-1}}\,$}
\newcommand{\parsec}{\texttt{PARSEC} }
\newcommand{\basti}{\texttt{BaSTI} }
\newcommand{\mist}{\texttt{MIST} }

\bibpunct{(}{)}{;}{a}{}{,} 
\newcommandtwoopt{\citeads}[3][][]{\href{http://adsabs.harvard.edu/abs/#3}%
	{\citealp[#1][#2]{#3}}} 
\newcommandtwoopt{\citepads}[3][][]{\href{http://adsabs.harvard.edu/abs/#3}%
	{\citep[#1][#2]{#3}}} 
\newcommandtwoopt{\citetads}[3][][]{\href{http://adsabs.harvard.edu/abs/#3}%
	{\citet[#1][#2]{#3}}}
\newcommandtwoopt{\citeyearads}[3][][]%
{\href{http://adsabs.harvard.edu/abs/#3}{\citeyear[#1][#2]{#3}}}

\usepackage{etoolbox}
\makeatletter
\patchcmd\@combinedblfloats{\box\@outputbox}{\unvbox\@outputbox}{}{%
	\errmessage{\noexpand\@combinedblfloats could not be patched}%
}%
\makeatother

\begin{document}
	
	\title{The Araucaria Project: High-precision orbital parallax and masses of eclipsing binaries from infrared interferometry\thanks{Based on observations made with ESO telescopes at Paranal and La Silla observatory under program IDs 087.C-0012(A), 087.C-0012(B), 089.C-0415(A), 089.C-0415(B), 092.C-0454(A), 093.C-0417(A), 094.C-0428(A), 094.D-0056(A), 096.C-0417(A), 096.D-0299(A), 097.D-0025(A), 097.C-0571(B), 098.C-0292(A), 0100.D-0024(A), 0100.D-0339(B).}}
	\titlerunning{High-precision orbital parallax and masses of eclipsing binaries}
	
	\subtitle{}
	\author{A.~Gallenne\inst{1,2},
		G.~Pietrzy\'nski\inst{3,4},
		D.~Graczyk\inst{4,5,6},
		B.~Pilecki\inst{3},
		J.~Storm\inst{7},
		N.Nardetto\inst{1},
		M.~Taormina\inst{3},
		W.~Gieren\inst{4},
		A.~Tkachenko\inst{8},
		P.~Kervella\inst{9},
		A.~M\'erand\inst{10} \and
		M.~Weber\inst{7}
	}
	
	\authorrunning{A. Gallenne et al.}
	
	\institute{European Southern Observatory, Alonso de C\'ordova 3107, Casilla 19001, Santiago, Chile
		\and Laboratoire Lagrange, UMR7293, Universit\'e de Nice Sophia-Antipolis, CNRS, Observatoire de la C\^ote dAzur, Nice, France
		\and Nicolaus Copernicus Astronomical Centre, Polish Academy of Sciences,  Bartycka 18, 00-716 Warszawa, Poland
		\and Universidad de Concepci\'on, Departamento de Astronom\'ia, Casilla 160-C, Concepci\'on, Chile
		\and Millenium Institute of Astrophysics, Av. Vicu{\~n}a Mackenna 4860, Santiago, Chile
		\and Centrum Astronomiczne im. Miko\l{}aja Kopernika, PAN, Rabia\'nska 8, 87-100 Toru\'n, Poland
		\and Leibniz-Institut f\"ur Astrophysik Potsdam (AIP), An der Sternwarte 16, 14482 Potsdam, Germany
		\and Instituut voor Sterrenkunde, KU Leuven, Celestijnenlaan 200D, B-3001 Leuven, Belgium
		\and LESIA, Observatoire de Paris, Universit\'e PSL, CNRS, Sorbonne Universit\'e, Univ. Paris Diderot, Sorbonne Paris Cit\'e, 5 place Jules Janssen, 92195 Meudon, France
		\and European Southern Observatory, Karl-Schwarzschild-Str. 2, 85748 Garching, Germany
	}

	\offprints{A. Gallenne} \mail{agallenn@eso.org}
	
	\date{Received May 06, 2019; accepted October 7, 2019}
	
	
	\abstract
	{The precise determinations of stellar mass at $\lesssim 1$\,\% provide important constraints on stellar evolution models. Accurate parallax measurements can also serve as independent benchmarks for the next Gaia data release.}
	{We aim at measuring the masses and distance of binary systems with a precision level better than 1\,\% using a fully geometrical and empirical method.}
	{We obtained the first interferometric observations for the eclipsing systems AI~Phe, AL~Dor, KW~Hya, NN~Del, $\psi$~Cen and V4090~Sgr with the VLTI/PIONIER combiner, which we combined with radial velocity measurements to derive their three-dimensional orbit, masses, and distance.}
	{We determined very precise stellar masses for all systems, ranging from 0.04\,\% to 3.3\,\% precision level. We combined these measurements with stellar effective temperature and linear radius to fit stellar isochrones models and determined the age of the systems. We also derived the distance to the systems with a precision level as high as 0.4\,\%.}
	{The comparison of theoretical models with stellar parameters shows that stellar models are still deficient in simultaneously fitting the stellar parameters ($T_\mathrm{eff}, R$ and $M$) with such level of precision on individual masses. This stresses the importance of precisely measuring the stellar parameters to better calibrate stellar evolution models. The precision of our model-independent orbital parallaxes varies from $24\mu$as as to 70$\mu$as
		and they provide a unique opportunity to check on the future Gaia measurements for possible systematic errors.}
	
	\keywords{techniques: interferometric -- techniques: high angular resolution -- stars: variables: Cepheids -- star: binaries: close}
	
	\maketitle
	
	%
	
	\defcitealias{Graczyk_2017_03_0}{Gr17}
	
	\newcommand{\diamAipheTwo}{0.161} 
	\newcommand{\ediamAipheTwo}{0.007} 
	\newcommand{\diamAipheOne}{0.102} 
	\newcommand{\ediamAipheOne}{0.004} 
	\newcommand{\fratioAiphe}{49.7} 
	\newcommand{\efratioAiphe}{2.4} 
	
	\newcommand{\diamAldor}{0.156} 
	\newcommand{\ediamAldor}{0.001} 
	\newcommand{\fratioAldor}{100.4} 
	\newcommand{\efratioAldor}{0.8} 

	\newcommand{\diamKwhyaOne}{0.228} 
	\newcommand{\ediamKwhyaOne}{0.008} 
	\newcommand{\diamKwhyaTwo}{0.159} 
	\newcommand{\ediamKwhyaTwo}{0.006} 
	\newcommand{\fratioKwhya}{45.8} 
	\newcommand{\efratioKwhya}{1.0} 
	
	\newcommand{\diamNndelOne}{0.115} 
	\newcommand{\ediamNndelOne}{0.001} 
	\newcommand{\diamNndelTwo}{0.084} 
	\newcommand{\ediamNndelTwo}{0.001} 
	\newcommand{\fratioNndel}{58.0} 
	\newcommand{\efratioNndel}{1.7} 
	
	\newcommand{\diamPsicenOne}{0.424} 
	\newcommand{\ediamPsicenOne}{0.007} 
	\newcommand{\diamPsicenTwo}{0.211} 
	\newcommand{\ediamPsicenTwo}{0.007} 
	\newcommand{\fratioPsicen}{17.1} 
	\newcommand{\efratioPsicen}{1.1} 
	
	\newcommand{\diamVsgrOne}{0.227} 
	\newcommand{\ediamVsgrOne}{0.030} 
	\newcommand{\diamVsgrTwo}{0.106} 
	\newcommand{\ediamVsgrTwo}{0.026} 
	\newcommand{\fratioVsgr}{0} 
	\newcommand{\efratioVsgr}{0} 
	
	\section{Introduction}
	
	In the course of the Araucaria project, different techniques for distance measurement are applied in order to track down the influence of the population effects on the most important standard candles like Cepheids, RR Lyrae stars, red clump stars, tip of the red-giant branch, etc... \citep{Gieren_2005_09_0,Gieren_2005_08_0}. Binary systems are of particular importance in our project, as demonstrated in our works with eclipsing binary systems, which have provided the most accurate distance of the Large and Small Magellanic Clouds \citep{Pietrzynski_2019_03_0,Pietrzynski_2013_03_0,Graczyk_2014_01_0}.
	
	Binary stars are the only tool enabling direct and accurate distance and stellar mass measurements. When they are double-lined spectroscopic binaries, the geometric distance can be measured, providing an independent benchmark to other measurements, such as for instance, the future Gaia parallax measurements \citep{Gaia-Collaboration_2016_11_0}. The mass is a fundamental parameter in order to understand the structure and evolution of stars, and precise measurements are necessary to check the consistency with theoretical models and tighter the constraints. For now, stellar parameters (e.g. the effective temperature, radius, ...) predicted from different stellar evolution codes can lead to discrepancies with the empirical values, and therefore providing a large range of possible age for a given system \citep[see e.g.][]{Torres_2010_02_0,Gallenne_2016_02_0}. Models of stellar interior differ in various ways, as for instance in the input physics, the initial chemical compositions, the treatment of convective-core overshooting, the rotational mixing or the mixing length parameter \citep{Marigo_2017_01_0,Bressan_2012_11_0,Dotter_2008_09_0,Pietrinferni_2004_09_0}. With high-precision measurements, evolutionary models can be tightly constrained and provide a better understanding of the stellar interior physics \citep{Higl_2018_09_0,Claret_2018_06_0,Valle_2017_04_0}.
	
	The precision on stellar parameters ($T_\mathrm{eff}, R, M, ...$) gradually improved with years thanks to eclipsing binary systems and ground- and space-based large photometric surveys (e.g. OGLE, WASP, KEPLER, LSST, TESS, ...). Combining radial velocities data with photometric observations during the eclipses, measurements at a $\sim 1-3$\,\% precision level are routinely achieved \citep[e.g.][]{Pilecki_2018_07_0,Pribulla_2018_08_0,Kirkby-Kent_2016_06_0,Graczyk_2015_09_0,Pilecki_2015_06_0,Gieren_2015_12_0,Pilecki_2013_12_0}, but recent work shows that precision on the stellar mass $<< 1$\,\% is necessary to obtain reliable determinations of the stellar interior model parameters \citep[overshooting, initial helium abundance, ...,][]{Higl_2018_09_0,Valle_2017_04_0}.
	
	Another model-independent approach for measuring stellar masses and geometrical distances with binary systems at $< 1$\,\% accuracy is to combine spectroscopic and astrometric observations \citep[see e.g.][]{Pribulla_2018_08_0,Gallenne_2016_02_0,Torres_2009_08_0, Zwahlen_2004_10_0,Morbey_1975_10_0,Herbison-Evans_1971__0}. This method does not require any assumptions. This has been recently applied to a Galactic binary Cepheid and has provided the most accurate distance and mass for a Milky Way Cepheid \citep{Gallenne_2018_11_0}. However, the systems need to be spatially resolved to enable astrometric measurements, which is not always the case with single-dish telescope observations where the components are too close. A higher angular resolution is provided from optical long-baseline interferometry (LBI), where close-in binary systems ($< 20$\,mas) can be detected. LBI already proved its efficiency in terms of angular resolution and accuracy for close-in binary stars \citep[see e.g.][]{Pribulla_2018_08_0,Gallenne_2018_11_0,Gallenne_2016_02_0,Gallenne_2015_07_0,Gallenne_2014_01_0,Gallenne_2013_04_0,Le-Bouquin_2013_03_0,Baron_2012_06_0}. 
	
	Recent work using LBI with eclipsing systems provided the most precise distances and masses for such systems, i.e. $< 0.4$\,\% and $< 0.1$\,\% respectively \citep{Pribulla_2018_08_0, Gallenne_2018_08_0,Gallenne_2016_02_0}. \citet{Valle_2017_04_0} demonstrated that such very precise mass determinations are required to obtain reliable determinations of stellar parameters such as the convective core overshooting, the initial helium abundance and the age of the system.
	
	In this paper, we report new observations with LBI of six eclipsing systems. The first goal of this interferometric observing program is to investigate the calibration of surface brightness-color (SBC) relations based only on eclipsing binary stars \citep[see][]{Graczyk_2017_03_0}. This will be published in a forthcoming specific paper. Here we focus on the precise measurements of mass and distance. Our sample is presented in Sect.~\ref{section__our_sample}. Section~\ref{section__observations} contains details about our interferometric observations, the data reduction process, and the radial velocities data we used. Our model fitting method is explained in Sect.~\ref{section__fitting_method} and the results for all systems are detailed in Sect.~\ref{section__results_for_individual_systems}. The evolutionary status is discussed in Sect.~\ref{section__discussion}. In this Section, we also compare our derived orbital parallaxes with the parallaxes from the second Gaia data release (GDR2). We then conclude in Sect.~\ref{section__conclusion}.

	\section{Our sample}
	\label{section__our_sample}
	
	\paragraph{AI~Phoenicis:} This 24.6\,d binary system is composed of a sub-giant star (K0IV) eclipsed by a main-sequence companion (F7V). This double-lined eclipsing system was discovered by \citet{Strohmeier_1972_04_0}, and has been extensively studied in spectroscopy and photometry over years \citep[see e.g.][]{Imbert_1979_06_0,Andersen_1988_05_0,Heminiak_2009_12_0,Kanjanascul_2012_06_0,Sybilski_2018_07_0}. Recently, variations in the systemic velocity seem to indicate that there is a possible wider component in the system (M. Konacki, priv, comm.), however, no orbital parameters have been derived so far.
	
	\citet{Kirkby-Kent_2016_06_0} provided precise mass measurements ($\sim 0.3$\,\%) which already enabled to test stellar evolution models, to constrain the mixing length and the helium abundance, and to derive an age of $4.39\pm0.32$\,Gyr. The last results of \citet{Sybilski_2018_07_0} gave masses precise to $\sim 0.08$\,\%, but there is no additional information about the stellar evolution. A third visual component located at about 11\arcsec was thought to be also bound to the AI~Phe system, however, \citet{Kirkby-Kent_2016_06_0} concluded that it is not physically associated. 
	
	From spectroscopic and photometric observations, \citet{Andersen_1988_05_0} derived the parallax to be $6.17\pm0.23$\,mas, consistent with the $5.94\pm0.24$\,mas from \citet{Graczyk_2017_03_0} using the same method and the $5.98\pm0.31$\,mas from \citet{Stassun_2016_12_0} from absolute stellar luminosity and bolometric flux fitting. The value from Hipparcos \citep[][assuming a single star]{van-Leeuwen_2007_11_0} is $\sim 2\sigma$ away from the more recent estimate \citep{Graczyk_2017_03_0}, but the Gaia measurement is within $1\sigma$ with $5.8336\pm0.0262$\,mas \citep{Gaia-Collaboration_2018_08_0}.
	
	\paragraph{AL~Doradus:} This eclipsing system has two identical main-sequence stars orbiting each other with a period of $14.9$\,d \citep[G0V+F9V][]{Graczyk_2019_02_0}. It was identified to be eclipsing by \citet{Kazarovets_1999_01_0} and has been poorly monitored in the past. The orbital parameters where only recently derived by \citet{Graczyk_2019_02_0} combining new photometric and spectroscopic observations, but they did not provide estimates of the masses and the distance. The parallax from Gaia is $15.1292\pm0.0286$\,mas, which corresponds to a distance of $66.97\pm0.13$\,pc, while the Hipparcos measurement \citep{van-Leeuwen_2007_11_0} has a lower accuracy with $16.07\pm0.43$\,mas, but roughly agrees with each other.
	
	\paragraph{KW~Hydrae:} The primary star is a chemically peculiar post-main-sequence A-type star (A5m, with metallic lines), and discovered to be double-lined spectroscopic binary by \citet{Chauville_1975_04_0}. She also determined the orbital parameters, and measured a period of 7.75\,days and a mass ratio of 1.14. The eclipsing nature of the system was later revealed by \citet{Reipurth_1981_09_0}. The secondary is a main-sequence star of spectral type F0V \citep{Hillenbrand_2004_04_0}. From new light curves and spectroscopic observations, \citet{Andersen_1984_01_0} redetermined the orbital parameters and provided the absolute dimensions of the system. They estimated masses with a precision of $1.5$\,\% and the distance to the system to be $86\pm4$\,pc ($\pi = 11.63\pm0.54$). From the absolute stellar luminosity and bolometric flux, \citet{Stassun_2016_12_0} derived a parallax of $11.75\pm0.62$\,mas, consistent with Andersen's value and the Hipparcos estimate ($11.53\pm0.42$\,mas, assuming a single star). These values all agrees with the Gaia measurement of $11.6744\pm0.0891$\,mas.
	
	\paragraph{NN~Delphini:} This star was first detected as variable by \citet{Makarov_1994_11_0}, and later identified to be eclipsing by \citet{Kazarovets_1999_01_0}. The first estimate of the orbital period of 99.3\,days was derived by \citet{Gomez-Forrellad_2003__0} from photometric observations. Later works combining spectroscopy and photometry was performed by \citet{Griffin_2014_06_0} who derived the full orbital parameters and the mass of both components to a precision level of $\sim 1.7$\,\%. Recent works of \citep{Sybilski_2018_07_0} refined the masses to a precision of $\sim 0.9$\,\%, combining also spectroscopy and photometry. The parallax of this system was determined with the Hipparcos satellite to be $6.28\pm0.89$\,mas \citep{van-Leeuwen_2007_11_0}, and recently refined with Gaia $5.6393\pm0.0636$\,mas.
	
	\paragraph{$\psi$~Centauri:} This A0IV star was recently discovered to be a 38.8\,d eclipsing system by \citep{Bruntt_2006_09_0} from photometric measurements. They inferred the effective temperature of both components which, combined with isochrone models, provided the approximate mass $M_1 = 3.1\pm0.3\,M_\odot$ and $M_2 = 2.0\pm0.2\,M_\odot$. Spectroscopic measurements were later obtained by \citet{Mantegazza_2010_03_0} and they disentangled the spectra of the two components. They independently estimated the spectroscopic orbital solutions, and combined them with the parameters derived from the light curve by \citep{Bruntt_2006_09_0} to determined the physical parameters of this system, such as the mass $M_1 = 3.084\pm0.016\,M_\odot$ and $M_2 = 1.891\pm0.030\,M_\odot$. These parameters were not updated since then.
	
	The parallax $\pi = 12.60\pm0.20$\,mas was provided by Hipparcos mission \citep[][assuming a single star]{van-Leeuwen_2007_11_0}, consistent with the prediction of \citet[$12.99\pm1.35$\,mas][]{Stassun_2016_12_0} from fitting the absolute stellar luminosity and bolometric flux. The Gaia DR2 gives a parallax $14.458\pm0.467$\,mas which is at $3.7\sigma$ from the Hipparcos value, probably for the same reasons as listed before.
	
	\paragraph{V4090~Sagittarii:} The eclipsing nature of this system was discovered by \citet{Waelkens_1983_04_0} from photometric light curve measurements. They reported a likely A5m spectral type for the primary and derived an orbital period of 11.415\,days. The spectral type was independently confirmed from spectroscopic lines to be between A1mA6-F0 \citep{Houk_1982__0}, which is also in agreement with the earlier work of \citet{Andersen_1977_09_0} who derived an A2m spectral type. The first photometric orbital solution of this eclipsing system were determined by \citet{Giuricin_1984_06_0} by analysing the data of \citet{Waelkens_1983_04_0}. They determined an early G spectral type for the secondary star and concluded that both components are on the main sequence. They also inferred $M_1 = 2.1\,M_\odot$ and $M_2 = 1.0\,M_\odot$ from their spectral type. \citet{North_1997_08_0} performed a more complete light curve analysis by combining  additional observations in different photometric bands. They combined them with the few available RVs from the literature \citep{Nordstrom_1985_07_0,Catchpole_1982__0} to update the physical properties of the system. Using evolutionary tracks, they derived $M_1 = 1.81\pm0.18\,M_\odot$ and inferred $M_2 = 0.95\pm0.08\,M_\odot$ from the mass function. They also estimated a distance of $85 \pm 10$\,pc, in agreement with Hipparcos ($84.5 \pm 7.0$\,pc) and Gaia ($92.46\pm0.71$\,pc).

	\section{Observations}
	\label{section__observations}
		
	\subsection{Astrometry}
	\label{section__interferometric_observations}
	
	Astrometric measurements were performed using near-IR long-baseline interferometry (LBI). We used the Very Large Telescope Interferometer \citep[VLTI ;][]{Woillez_2018_07_0} with the four-telescope combiner PIONIER \citep{Le-Bouquin_2011_11_0} to measure the squared visibilities and the closure phases. PIONIER combines the light coming from four telescopes in the $H$ band, either in a broadband mode or with a low spectral resolution, where the light is dispersed into six spectral channels. The recombination from all four telescopes simultaneously provides six visibility and four closure phase measurements across a range of spectral channels.
	
	Our observations were carried out from 2016 to 2017 using the 1.8\,m Auxiliary Telescopes with various configurations, providing six projected baselines ranging from 40 to 140\,m. Data were dispersed over six spectral channels across the $H$ band ($1.50-1.80\,\mathrm{\mu m}$). To monitor the instrumental and atmospheric contributions, the standard observational procedure, which consists of interleaving the science target by reference stars, was used. The calibrators, listed in Table~\ref{table__astrometry} and detailed in Table~\ref{table_calibrators}, were selected using the \textit{SearchCal}\footnote{Available at \url{http://www.jmmc.fr/searchcal}.} software \citep{Bonneau_2006_09_0,Bonneau_2011_11_0} provided by the Jean-Marie Mariotti Center (JMMC).
	
\begin{figure*}[!h]
		\centering
		\resizebox{\hsize}{!}{\includegraphics{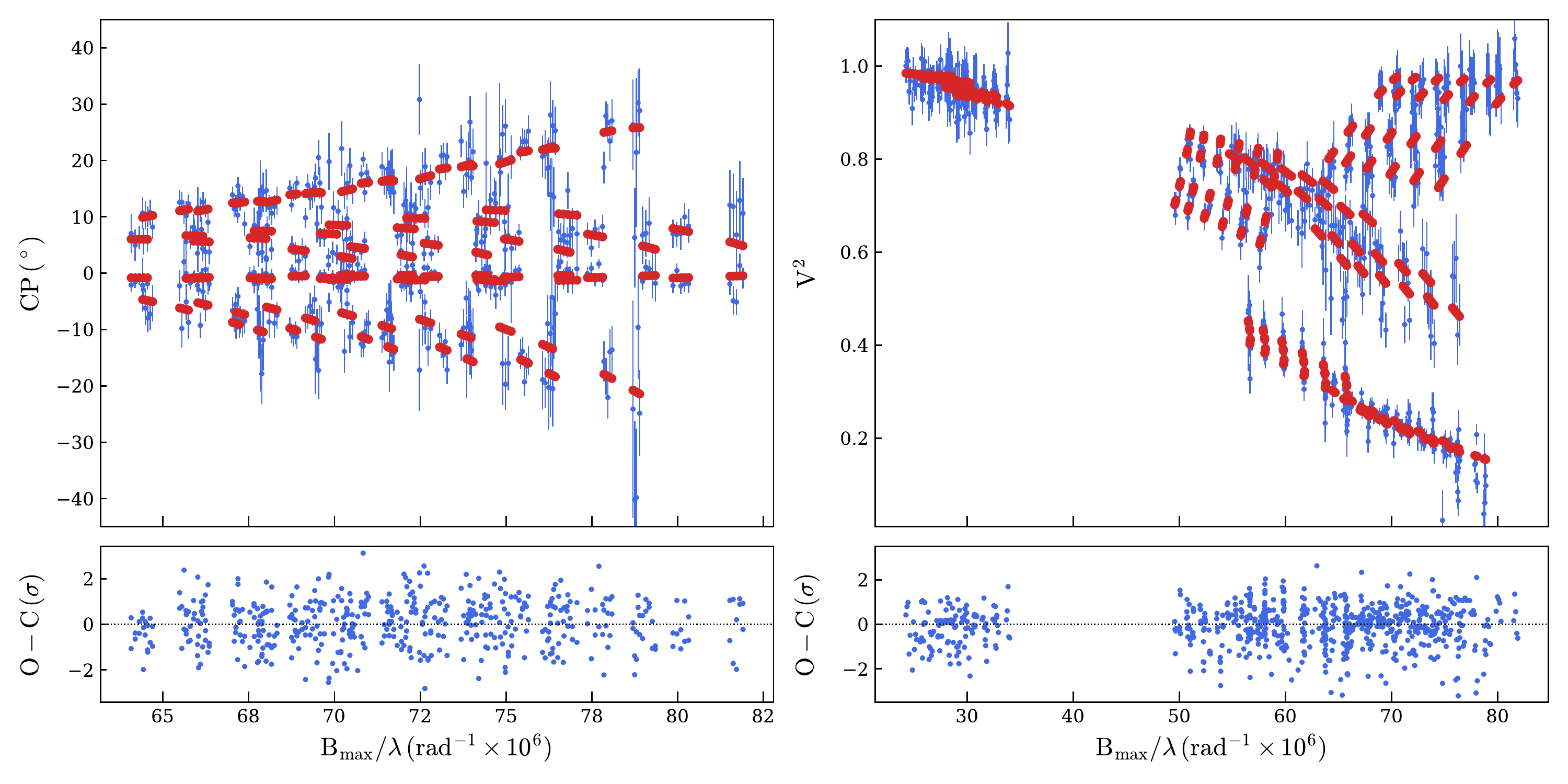}}
		\caption{Closure phase and squared visibility measurements of the last measurements of AI~Phe. The data are in blue, while the red dots represent the fitted binary model for this epoch. The residuals (in number of sigmas) are also shown on the bottom panels.}
		\label{figure__visibility_aiphe}
	\end{figure*}

	The data have been reduced with the \textit{pndrs} package described in \citet{Le-Bouquin_2011_11_0}. The main procedure is to compute squared visibilities and triple products for each baseline and spectral channel, and to correct for photon and readout noises. The calibrated data are available at the Optical Interferometry DataBase\footnote{\url{http://oidb.jmmc.fr/index.html}.} In Fig.~\ref{figure__visibility_aiphe}, an example of the squared visibilities and closure phases for the last observation of AI~Phe are presented, and where the binary nature of the system is clearly detected.
	
	%
	For each epoch, we proceeded to a grid search to find the global minimum and the location of the companion. For this we used the interferometric tool \texttt{CANDID}\footnote{Available at \url{https://github.com/amerand/CANDID}} \citep{Gallenne_2015_07_0} to search for the companion using all available observables. \texttt{CANDID} allows a systematic search for point-source companions performing a $N \times N$ grid of fit, whose minimum needed grid resolution is estimated a posteriori. The tool delivers the binary parameters, namely the flux ratio $f$ and the relative astrometric separation ($\Delta \alpha, \Delta \delta$). \texttt{CANDID} can also fit the angular diameter of both components, however, in our cases, we kept them fixed during the fitting process as the VLTI baselines do not allow reliable measurements of such small diameters. For each epoch, \texttt{CANDID} find the global best fit separation vector. The final astrometric positions for all epochs of all systems are listed in Table~\ref{table__astrometry}. We estimated the uncertainties from the bootstrapping technique (with replacement) and 10~000 bootstrap samples (also included in the \texttt{CANDID} tool). For the flux ratio, we took from the distributions the median value and the maximum value between the 16th and 84th percentiles as uncertainty. For the astrometry, the $1\sigma$ error region of each position $(\Delta \alpha, \Delta \delta)$ is defined with an error ellipse parametrized with the semi-major axis $\sigma_\mathrm{maj}$, the semi-minor axis $\sigma_\mathrm{min}$, and the position angle $\sigma_\mathrm{PA}$ measured from north through east. We also quadratically added to ($\sigma_\mathrm{maj}, \sigma_\mathrm{min}$) the systematic uncertainty of 0.35\,\% from the precision of the PIONIER wavelength calibration \citep{Gallenne_2018_08_0,Kervella_2017_01_1}. This systematic error is one of the fundamental limitation of our analysis: it limits the final accuracy on the apparent semi-major axis in the orbital fit ("$a$ (mas)" in table~\ref{table__orbits}). The distance being inversely proportional to apparent semi-major axis, the accuracy on the distance is limited also by the same amount of 0.35\,\%.
	
	As mentioned, the angular diameter of some of the stars are too small to be spatially resolved by the VLTI. So we kept them fixed during the grid search. For AI~Phe, we fixed the values to $\theta_\mathrm{LD_1} = \diamAipheOne \pm\ediamAipheOne$,mas and $\theta_\mathrm{LD_2} =  \diamAipheTwo \pm \ediamAipheTwo$,mas, as estimated by  \citet[][hereafter \citetalias{Graczyk_2017_03_0}]{Graczyk_2017_03_0}. For AL Dor, we adopted the values $\theta_\mathrm{LD_1} = \theta_\mathrm{LD_2} = \diamAldor \pm\ediamAldor$\,mas from \citet{Graczyk_2019_02_0}. We adopted $\theta_\mathrm{LD_1} = \diamKwhyaOne \pm\ediamKwhyaOne$\,mas and $\theta_\mathrm{LD_2} =  \diamKwhyaTwo \pm \ediamKwhyaTwo$\,mas from \citetalias{Graczyk_2017_03_0} for KW~Hya. For $\psi$~Cen we used $\theta_\mathrm{LD_1} =  \diamPsicenOne \pm \ediamPsicenOne$\,mas and $\theta_\mathrm{LD_2} =  \diamPsicenTwo \pm \ediamPsicenTwo$\,mas \citep{Mantegazza_2010_03_0}. We adopted $\theta_\mathrm{LD_1} =  \diamVsgrOne \pm \ediamVsgrOne$\,mas and $\theta_\mathrm{LD_2} =  \diamVsgrTwo \pm \ediamVsgrTwo$\.mas from \citet{North_1997_08_0} for V4090~Sgr. Finally, for NN~Del we used $\theta_\mathrm{LD_1} =  \diamNndelOne \pm \ediamNndelOne$\,mas and $\theta_\mathrm{LD_2} =  \diamNndelTwo \pm \ediamNndelTwo$\,mas from \citet{Sybilski_2018_07_0}. Except AI~Phe and KW~Hya, the angular diameters were determined from the combination of the linear radius and the Gaia parallax. We note that angular diameters have no impact on the measured astrometry.
	
\begin{figure}[!h]
		\centering
		\resizebox{\hsize}{!}{\includegraphics{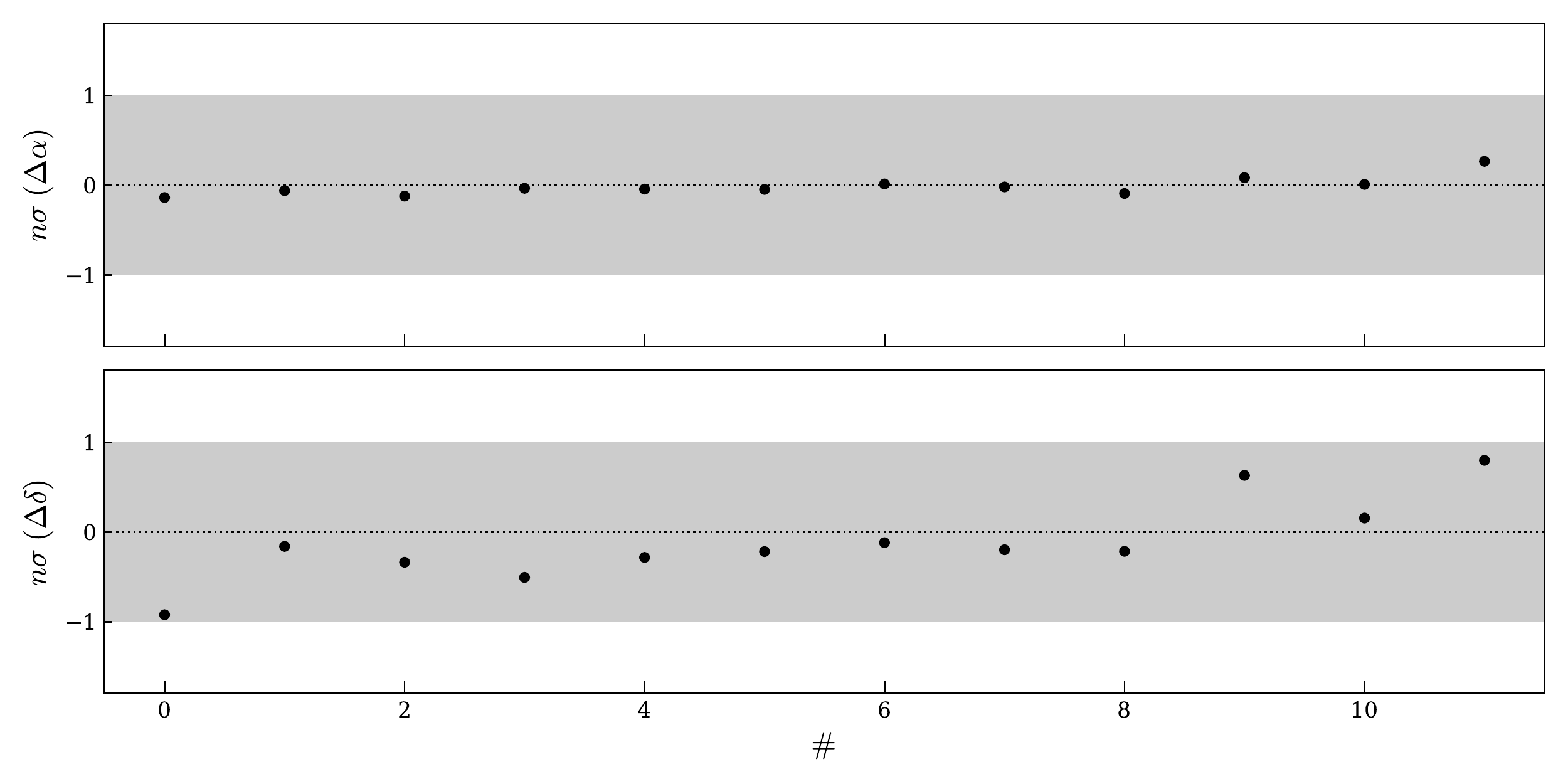}}
		\caption{Difference in number of sigma of the astrometric positions $(\Delta \alpha, \Delta \delta)$ for a fitted and fixed flux ratio. This is calculated as $(\Delta X_\mathrm{fitted\ f} - \Delta X_\mathrm{fixed\ f})/\sqrt{\sigma_\mathrm{X, fitted\ f}^2 + \sigma_\mathrm{X, fixed\ f}^2}$ with $X$ being $\Delta \alpha$ or $\Delta \delta$. }
		\label{figure__comparison_fixed_f}
	\end{figure}

	Finally, we also investigated the impact of fitting or fixing the flux ratio in deriving our astrometric positions for very nearby components (i.e. $<\lambda/2B$). For this we used the star AI~Phe which has a component with the closest orbit. We repeated the process with \texttt{CANDID} (i.e. the grid search and then the bootstrapping) by keeping the flux ratio to the average value $f = 49.7$\,\%. The differences for our 12 astrometric positions $(\Delta \alpha, \Delta \delta)$ with and without fitting the flux ratio are displayed in Fig.~\ref{figure__comparison_fixed_f}. We notice that there is no significant impact of removing $f$ from the fitted parameters as the agreement within $1\sigma$.

	\begin{table*}[!ht]
		\centering
		\caption{Relative astrometric position of the secondary component for all systems.}
		\begin{tabular}{ccccccccc} 
			\hline
			\hline
			HJD &  $\Delta \alpha$	&  $\Delta \delta$	& $\sigma_\mathrm{PA}$	& $\sigma_\mathrm{maj}$	&  $\sigma_\mathrm{min}$	&  $f$	&  Baselines & Calibrators	\\
			&  (mas)			& (mas)			& ($^\circ$)	& (mas)		&  (mas)  &  (\%)  & 	&	\#  \\
			\hline
			\multicolumn{9}{c}{AI~Phe} \\
			2457717.594 & 0.334 & 1.358 & 71.3 & 0.021 & 0.011 &  $56.2\pm2.6$  &  D0-G2-J3-K0   &   1,2,3 \\
			2457752.579 & -0.258 & -1.171 & -43.7 & 0.018 & 0.014 &  $48.9\pm2.1$  &  A0-G1-J2-J3  & 1,2,3 \\
			2457754.608 & -0.200 & -1.071 & -67.8 & 0.035 & 0.014 &  $48.0\pm6.0$  &  A0-G1-J2-J3  & 1,2,3 \\
			2458022.705 & -0.276 & -1.151 & 93.3 & 0.032 & 0.014 &  $46.9\pm3.7$  &  A0-G1-J2-J3  &  4,5,6 \\
			2458032.807 & 0.203 & 0.767 & -11.5 & 0.016 & 0.015 &  $52.3\pm4.5$  &  A0-G1-J2-J3  &   4,5,6 \\
			2458034.680 & 0.298 & 1.127 & 51.9 & 0.008 & 0.008 &  $50.1\pm1.0$ &  A0-G1-J2-J3  & 4,5,6 \\
			2458035.669 & 0.312 & 1.256 & 19.8 & 0.005 & 0.006 &  $49.8\pm0.5$  &  A0-G1-J2-J3  &    4,5,6 \\
			2458046.662 & -0.268 & -1.057 & 62.3 & 0.015 & 0.011 &  $49.2\pm2.3$  &  D0-G2-J3-K0  &  4,5,6 \\
			2458047.752 & -0.273 & -1.177 & 64.2 & 0.026 & 0.009 &  $48.0\pm1.9$  &  A0-B2-D0-J3  &  4,5,6 \\
			2458060.630 & 0.312 & 1.293 & -34.6 & 0.011 & 0.009 &  $48.6\pm0.8$  &  A0-G1-J2-J3  &   4,5,6 \\
			2458061.643 & 0.332 & 1.339 & -6.3 & 0.008 & 0.007 &  $49.5\pm0.5$  &  A0-G1-J2-J3  &    4,5,6 \\
			2458083.661 & 0.282 & 1.103 & 87.5 & 0.006 & 0.006 &  $48.5\pm0.9$  &  A0-G1-J2-J3  &    4,5,6 \\
			\hline
			\multicolumn{9}{c}{AL~Dor} \\
			2457387.578 & -1.471 & -1.520 & 99.0 & 0.033 & 0.022   &  $99.9\pm0.7$  &  A0-G1-J2-J3  &  1,2,3  \\
			2457418.579 & -1.399 & -1.491 & 102.7 & 0.016 & 0.009    &  $100.0\pm0.6$  &  A0-G1-J2-J3  &  1,2,3  \\
			2457678.758 & 1.617 & 1.667 & 73.1 & 0.019 & 0.010 &  $100.1\pm0.3$  &  A0-G1-J2-J3  &  1,2,3  \\
			2457752.668 & 1.477 & 1.511 & 108.7 & 0.017 & 0.008  &  $99.5\pm1.3$  &  A0-G1-J2-J3  &  1,2,3  \\
			2457753.626 & 1.679 & 1.715 & 107.7 & 0.014 & 0.009  &  $99.8\pm0.6$  &  A0-G1-J2-J3  &  1,2,3  \\
			2457780.581 & 0.727 & 0.635 & -32.1 & 0.014 & 0.011  &  $102.7\pm5.4$  &  A0-G1-J2-J3  &  1,2,3  \\
			2457826.532 & 1.260 & 1.263 & 10.8 & 0.010 & 0.007 &  $100.1\pm0.4$  &  A0-G1-J2-J3  &  1,2,3  \\
			2458034.790 & 1.085 & 1.070 & 46.8 & 0.010 & 0.008 &  $100.3\pm0.8$  &  A0-G1-J2-J3  &  2,3,4  \\
			2458035.793 & 1.476 & 1.465 & -5.4 & 0.010 & 0.009 &  $100.0\pm0.3$  &  A0-G1-J2-J3  &  2,3,4  \\
			2458036.798 & 1.678 & 1.728 & 75.7 & 0.020 & 0.010 &  $99.9\pm0.5$  &  A0-G1-J2-K0  &  2,3,4  \\
			2458044.751 & -1.353 & -1.458 & 49.5 & 0.018 & 0.006   &  $100.3\pm0.3$  &  A0-G1-J2-K0  &  2,3,4  \\
			2458045.751 & -0.970 & -1.108 & 50.1 & 0.048 & 0.011   &  $100.3\pm2.3$  &  A0-G1-J2-K0  &  2,3,4  \\
			2458058.767 & -1.511 & -1.573 & 107.2 & 0.017 & 0.009    &  $101.3\pm0.5$  &  A0-G1-J2-J3  &  2,3,4  \\
			2458080.657 & 1.516 & 1.529 & 89.3 & 0.011 & 0.008 &  $100.0\pm0.2$  &  A0-G1-J2-J3  &  2,3,4  \\
			2458081.765 & 1.694 & 1.748 & 86.1 & 0.008 & 0.008 &  $99.7\pm0.2$  &  A0-G1-J2-J3  &  2,3,4  \\
			2458083.761 & 1.244 & 1.325 & -50.5 & 0.014 & 0.007  &  $100.0\pm0.4$  &  A0-G1-J2-J3  &  2,3,4  \\
			2458116.580 & -0.964 & -0.949 & 103.4 & 0.014 & 0.006    &  $100.7\pm0.8$  &  A0-G1-J2-J3  &  2,3,4  \\
			2458117.675 & -1.418 & -1.479 & 91.3 & 0.008 & 0.006   &  $101.2\pm0.4$  &  A0-G1-J2-K0  &  2,3,4  \\
			2458118.711 & -1.474 & -1.549 & 113.8 & 0.011 & 0.008    &  $101.5\pm0.8$  &  A0-G1-J2-K0  &  2,3,4  \\
			\hline
			\multicolumn{9}{c}{KW~Hya} \\
			2457481.618 & 1.232 & 0.284 & 87.5 & 0.006 & 0.003 &  $45.8\pm0.3$  &  A0-G1-J2-J3  &  1,2,3  \\
			2457482.575 & 1.332 & 0.365 & 101.3 & 0.008 & 0.003 &  $46.0\pm0.2$  &  A0-G1-J2-J3  &  1,2,3  \\
			2457512.534 & 1.192 & 0.274 & 114.5 & 0.008 & 0.006 &  $45.2\pm0.7$  &  A0-G1-J2-J3  &  1,2,3  \\
			2457750.837 & -0.694 & -0.228 & -25.4 & 0.024 & 0.013 &  $44.2\pm5.6$  &  A0-G1-J2-J3  &  1,2,3  \\
			2457753.860 & 1.319 & 0.354 & -53.7 & 0.007 & 0.004 &  $45.8\pm0.3$  &  A0-G1-J2-J3  &  1,2,3  \\
			2457754.845 & 0.769 & 0.232 & -34.4 & 0.010 & 0.005 &  $47.7\pm2.5$  &  A0-G1-J2-J3  &  1,2,3  \\
			\hline
			\multicolumn{9}{c}{NN~Del} \\
			2457954.758 & 0.094 & 3.588 & 104.2 & 0.012 & 0.016  &  $59.9\pm1.8$  &  A0-G1-J2-J3  & 1,2,3  \\
			2457994.694 & 0.027 & 1.257 & 14.5 & 0.007 & 0.006 &  $57.4\pm0.7$  &  A0-G1-J2-J3  &  1,2,3  \\
			2458002.682 & 0.061 & 2.830 & 74.4 & 0.008 & 0.012 &  $55.6\pm1.5$  &  A0-G1-J2-J3  &  1,2,3  \\
			2458016.615 & 0.113 & 4.637 & 78.8 & 0.008 & 0.017 &  $59.0\pm0.4$  &  A0-G1-J2-J3  &  1,2,3  \\
			\hline
			\multicolumn{9}{c}{$\psi$~Cen} \\
			2457481.781 & -1.670 & 0.896 & -54.0 & 0.014 & 0.010 &  $17.7\pm0.2$  &  A0-G1-J2-J3  &  1,2,3  \\
			2457511.768 & -5.941 & 2.909 & -10.3 & 0.030 & 0.013 &  $17.4\pm0.5$  &  A0-G1-J2-J3  &  1,2,3  \\
			2457604.528 & 2.093 & -1.001 & 18.2 & 0.013 & 0.008  &  $17.9\pm0.3$  &  A0-G1-J2-J3  &  1,2,3  \\
			2457622.494 & -6.862 & 3.327 & -17.8 & 0.029 & 0.016 &  $18.2\pm0.3$  &  A0-G1-J2-J3  &  1,2,3  \\
			2457779.885 & -6.643 & 3.267 & 66.2 & 0.029 & 0.014  &  $16.6\pm0.6$  &  A0-G1-J2-J3  &  1,2,3  \\
			2457816.805 & -6.804 & 3.321 & -17.6 & 0.033 & 0.022 &  $14.9\pm0.5$  &  A0-G1-J2-J3  &  1,2,3  \\
			\hline
			\multicolumn{9}{c}{V4090~Sgr} \\
			2457511.852 & -0.670 & -1.276 & 100.4 & 0.020 & 0.012  &  $14.5\pm0.3$  &  A0-G1-J2-J3  &  1,2,3  \\
			2457512.906 & -0.814 & -1.385 & -2.9 & 0.022 & 0.016  &  $13.4\pm0.3$  &  A0-G1-J2-J3  &  1,2,3  \\
			2457597.769 & 0.750 & 1.334 & -51.7 & 0.032 & 0.014  &  $12.7\pm0.2$  &  A0-G1-J2-J3  &  1,2,3  \\
			2457599.572 & 0.612 & 1.031 & -56.6 & 0.017 & 0.009  &  $13.3\pm0.4$  &  A0-G1-J2-J3  &  1,2,3  \\
			2457602.803 & -0.626 & -1.217 & -26.6 & 0.060 & 0.027  &  $13.1\pm0.7$  &  A0-G1-J2-J3  &  1,2,3  \\
			2457684.583 & -0.750 & -1.252 & -13.6 & 0.085 & 0.041  &  $14.2\pm0.9$  &  A0-G1-J2-J3  &  1,2,3 \\
			\hline
		\end{tabular}
		\label{table__astrometry}
	\end{table*}
	
	\subsection{Radial velocities}	

	\paragraph{AI~Phe:} We used radial velocity measurements from \citet{Sybilski_2018_07_0} and \citet{Heminiak_2009_12_0} only, which are the most precise. In addition, we retrieved 33 HARPS spectra \citep{Mayor_2003_12_0} from the ESO archive. Thirty spectra were taken in full resolution mode, and three in EGGS mode, covering the spectral range 3900-6900\,\AA. HARPS data were processed with the standard ESO/HARPS pipeline reduction package. To extract the RVs we used the Broadening Function (BF) formalism \citep{Rucinski_1992_11_0} implemented in the RaveSpan software\footnote{\url{https://users.camk.edu.pl/pilecki/ravespan/}} \citep[][see also e.g. \citealt{Pilecki_2018_07_0, Gallenne_2016_02_0, Graczyk_2015_09_0,Pilecki_2013_12_0}]{Pilecki_2017_06_0}. The measurements errors were estimated using the uncertainties of the broadening profiles fits.  The velocities are listed in Table~\ref{table_rv_aiphe}.
	
	\paragraph{AL~Dor:} We used the radial velocity measurements from  \citet{Graczyk_2019_02_0}. There are no other data available in the literature.

	\paragraph{KW~Hya:} For our analysis, we used the radial velocity measurements of \citet{Andersen_1984_01_0}.

	\paragraph{NN~Del:} We collected the velocity measurements from \citet{Griffin_2014_06_0} and \citet{Sybilski_2018_07_0}. We also completed the sample of RVs with new observations from the STELLA echelle spectrograph \citep[SES,][]{Strassmeier_2010__0}. It is mounted on the robotic 1.2\,m STELLA-II telescope in Tenerife, Spain and provides high-resolution spectra in the wavelength range 3900-8700\,\AA\, ($R \sim 55~000$). The spectra were reduced using the STELLA data-reduction pipeline \citep{Weber_2008_08_0}. To extract the RVs we used the RaveSpan software as explained previously for the AI~Phe system.  The velocities are listed in Table~\ref{table_rv_nndel}.
	
	\paragraph{$\psi$~Cen:} The only spectroscopic observations of this system in the literature are the ones from \citet{Mantegazza_2010_03_0}, but in the method they used (spectral disentangling) individual radial velocities are not calculated and thus are not provided. As we need them for our analysis, we collected their reduced spectra (kindly provided by M. Rainer) and extracted RVs using our own method. These data were supplemented by our own spectra taken with the HARPS spectrograph \citep{Mayor_2003_12_0} in a much recent epoch. Spectra were reduced using the ESO data-reduction pipeline. To extract the RVs we used the RaveSpan software as explained previously for the AI~Phe system. The use of BF method was crucial to obtain reliable RV measurements for $\psi$~Cen because of very high rotational broadening of the lines and blending of the profiles, especially at phases where the separation of orbital velocities are small. If individual RVs were not needed, the only other method that could give similar results in this case would be the spectral disentangling as used by \citet{Mantegazza_2010_03_0}. The velocities are listed in Table~\ref{table_rv_psicen}.
	
	\paragraph{V4090~Sgr:} There are unfortunately no velocity measurements of the secondary component as spectral disentangling is difficult. We used the only three RVs of the primary estimated by \citep{Nordstrom_1985_07_0}. It exists apparently six other velocities from \citet{Catchpole_1982__0} but as far as we know they are not available online unfortunately.

	\section{Fitting method}
	\label{section__fitting_method}
	
	To determine the best fit parameters we simultaneously fit the radial velocities and astrometric positions using a Markov Chain Monte Carlo routine\footnote{With the Python package \texttt{emcee} developed by \citet{Foreman-Mackey_2013_03_0}.}, whose the log-likelihood function is given as
	\begin{displaymath}
	\log(\mathcal{L}) = - \dfrac{1}{2}\,\chi^2,\, \mathrm{with}\,\chi^2 = \chi^2_\mathrm{RV} + \chi^2_\mathrm{ast}
	\end{displaymath}
	
	$\chi^2_\mathrm{RV}$ defines the radial velocity measurements with
	\begin{displaymath}
	\chi^2_\mathrm{RV} =  \sum \dfrac{(V_1 - V_\mathrm{1m})^2}{\sigma_\mathrm{V_1}^2} +  \sum \dfrac{(V_2 - V_\mathrm{2m})^2}{\sigma_\mathrm{V_2}^2}, 
	\end{displaymath}
	in which $V_\mathrm{i}$ and $\sigma_\mathrm{V_i}$ denotes the measured radial velocities and uncertainties for the component $i$. Errorbars are often underestimated in the literature, so all of them were rescaled according to the average scatter of the fit. ($V_\mathrm{1m}, V_\mathrm{2m})$ are the Keplerian velocity models of both components, defined by \citep{Heintz_1978__0}
	\begin{eqnarray*}
		V_\mathrm{1m} &=& \gamma + K_1\,[\cos(\omega + \nu) + e\cos{\omega}], \\
		V_\mathrm{2m} &=& \gamma + K_2\,[\cos(\omega + \nu) + e\cos{\omega}], \\
		\tan \dfrac{\nu}{2} &= & \sqrt{\dfrac{1 + e}{1 - e}} \tan \dfrac{E}{2}, \\
		E - e \sin E &=& \dfrac{2\pi (t - T_\mathrm{p})}{P_\mathrm{orb}}
	\end{eqnarray*}
	where $\gamma$ is the systemic velocity, $e$ the eccentricity, $\omega$ the argument of periastron, $\nu$ the true anomaly,  $E$ the eccentric anomaly, $t$ the observing date, $P_\mathrm{orb}$ the orbital period, and $T_\mathrm{p}$ the time of periastron passage. The parameters $K_1$ and  $K_2$ are the radial velocity amplitude of both stars. 
	
	$\chi^2_\mathrm{astro}$ defines the astrometric measurements such as
	\begin{eqnarray*}
		&\chi^2_\mathrm{ast} =& \chi^2_\mathrm{a} + \chi^2_\mathrm{b}, \\
		&\chi^2_\mathrm{a} =& \sum \frac{[ (\Delta \alpha - \Delta \alpha_\mathrm{m}) \sin \sigma_\mathrm{PA} + (\Delta \delta - \Delta \delta_\mathrm{m}) \cos \sigma_\mathrm{PA} ]^2}{\sigma^2_\mathrm{maj}}, \\
		&\chi^2_\mathrm{b} =& \sum \frac{[ -(\Delta \alpha - \Delta \alpha_\mathrm{m}) \cos \sigma_\mathrm{PA} + (\Delta \delta - \Delta \delta_\mathrm{m}) \sin \sigma_\mathrm{PA} ]^2}{\sigma^2_\mathrm{min}}, \\
	\end{eqnarray*}
	in which $(\Delta \alpha, \Delta \delta, \sigma_\mathrm{PA}, \sigma_\mathrm{maj}, \sigma_\mathrm{min})$ denote the relative astrometric measurements with the corresponding error ellipses, and $(\Delta \alpha_\mathrm{m}, \Delta \delta_\mathrm{m})$ the astrometric model defined with:
	\begin{eqnarray*}
		\Delta \alpha_\mathrm{m} &=& r \,[ \sin \Omega \cos(\omega + \nu) + \cos i \cos \Omega \sin(\omega + \nu) ], \\
		\Delta \delta_\mathrm{m} &=& r \,[ \cos \Omega \cos(\omega + \nu) - \cos i \sin \Omega \sin(\omega + \nu) ], \\
		r &=& \dfrac{a (1 - e^2)}{1 + e\cos \nu},\\
	\end{eqnarray*}
	where $\Omega$ is the longitude of ascending node, $i$ the orbital inclination, and $a$ the angular semi-major axis.
	
	
	As a starting point for our 100 MCMC walkers, we performed a least squares fit using orbital values from the literature as first guesses. We then ran 100 initialization steps to well explore the parameter space and get settled into a stationary distribution. For all cases, the chain converged before 50 steps. Finally, we used the last position of the walkers to generate our full production run of 1000 steps, discarding the initial 50 steps. All the orbital elements, i.e. $P_\mathrm{orb}, T_\mathrm{p}, e, \omega, \Omega, K_1, K_2, \gamma, a$ and $i$, are estimated from the distribution taking the median value and the maximum value between the 16th and 84th percentiles as uncertainty (although the distributions were roughly symmetrical).
	
	From these distributions, we derive the distributions of the mass of both components and the distance to the system with \citep{Torres_2010_02_0}
	\begin{eqnarray*}
		M_1 &=& \dfrac{1.036149\times 10^{-7} (K_1 + K_2)^2 K_2\,P\,(1 - e^2)^{3/2} }{\sin ^3 i},\\
		M_2 &=& \dfrac{1.036149\times 10^{-7} (K_1 + K_2)^2 K_1\,P\,(1 - e^2)^{3/2} }{\sin ^3 i},\\
		a_\mathrm{AU} &=& \dfrac{9.191940\times 10^{-5} (K_1 + K_2)\,P \sqrt{1 - e^2} }{\sin i}, \\
		d &=& \dfrac{a_\mathrm{AU}}{a}
	\end{eqnarray*}
	where the masses are expressed in solar units, the distance in parsec, $K_1$ and $K_2$ in \kms, $P$ in days, and $a$ in arcsecond. The parameter $a_\mathrm{AU}$ is the linear semi-major axis expressed in astronomical units (the constant value of \citet{Torres_2010_02_0} is expressed in solar radii, and was converted using the astronomical constants $R_\odot = 695.658 \pm 0.140 \times 10^6$\,m from \citealt{Haberreiter_2008_03_0} and $AU = 149~597~870~700 \pm 3$\,m from \citealt{Pitjeva_2009_04_0}). As previously, we then took the median value and the maximum value between the 16th and 84th percentiles as uncertainty. The fitting results are presented in the next section for all systems. Note that for the angular size of the orbit (and so the distance), we additionally took into account the systematic uncertainty from the interferometric wavelength calibration, i.e. we also quadratically added 0.35\,\% to the final uncertainty.
	
	\section{Results for individual systems}
	\label{section__results_for_individual_systems}
	
	\subsection{AI~Phoenicis}
	\label{subsection__results_aiphe}
	
	\begin{figure*}[!h]
		\centering
		\resizebox{\hsize}{!}{\includegraphics{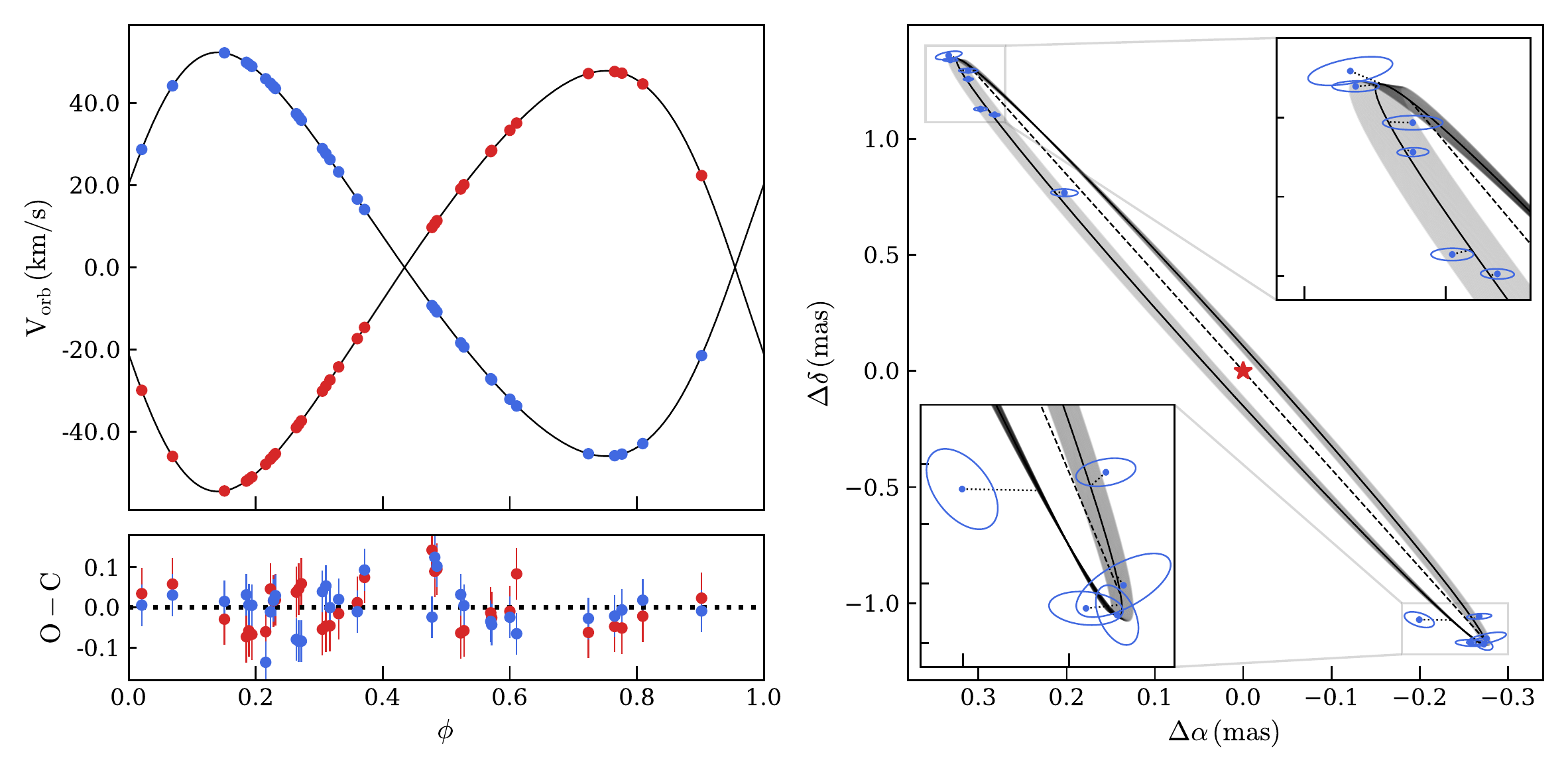}}
		\caption{\textit{Left}: radial velocities of the primary (red) and the secondary (blue) in the AI~Phe system. \textit{Right}: PIONIER astrometric orbit. The shaded grey area represents the $1\sigma$ orbit.}
		\label{figure__orbitaiphe}
	\end{figure*}
	
	\begin{table*}[!ht]
		\centering
		\caption{Best fit orbital elements and parameters for our binary systems.}
		\begin{tabular}{ccccccc}
			\hline
			\hline
			Parameter 	& AI~Phe & AL~Dor  	&  KW~Hya & NN~Del & $\psi$~Cen & V4090~Sgr	\\
			\hline
			$P_\mathrm{orb}$ (days)									& 24.59215(2) & 	 14.90537(1)		& 7.750468(6) & 99.2690(9)	&	38.8121(2)	&  11.41497(2)  	\\
			$T_\mathrm{p}$ (days)									& 8248.603(4) &		7764.791(2)	 &	24.421(32)	& 5235.164(70) & 8180.836(17) & 2442657.8(5) \\
			$e$																     & 	0.1872(1) &		 	0.1952(1)   &	0.094(4)	& 0.5197(4)	 & 0.550(1)	  & 0.012(6)\\
			$K_1$ (\kms)					  							& 	51.166(8) &	57.367(8)		&	70.14(18)	& 36.191(23) &	49.51(16) &		47.7(2) 	\\
			$K_2$ (\kms)					  							& 	49.118(7) &		57.420(9)	 	&	93.20(62)	& 39.407(37) & 80.48(33)	&	92(1)		 	\\
			$\gamma$	(\kms)											& 	$-$2.111(4) &	11.805(4)		&	$-$4.31(17)/$-$1.87(68)\,\tablefootmark{b} & $-$9.485(15)	& 	3.58(9) 	  & 	$-$36.3(1) \\
			$\omega$	($\degr$)									 & 	110.36(3) &		287.48(3)	  	&	225.38(1.6)	& 350.00(7)	& 19.45(27)	& 57(17) 	\\
			$\Omega$	($\degr$)									 &	13.31(23) &		223.83(7)	&	75.76(11)	& 181.41(14)	 &	115.79(10) & 29.0(3) \\
			$a$ (mas)													 & 	 1.315(5) &		 2.344(9)		&	 1.329(7)	& 3.508(13) & 5.055(20)	& 1.596(11) \\
			$a$ (AU)														 & 	 0.2227(8) &	 0.1543(5)		&	 0.1160(6)	& 0.5894(21) & 0.3874(17)  & 0.1471(14)	\\
			$i$ ($\degr$)													& 	91.32(39) &		91.21(11)	&	92.50(26)	& 89.90(11)	& 89.20(13)	& 87.4(6) \\
			\hline
			$M_1$ ($M_\odot$)							&	1.1941(7) &	1.1029(4)  &	1.975(28) & 1.4445(29) & 3.187(31)  & 2.15(7) \\
			$M_2$ ($M_\odot$)							& 	1.2438(7) &  	1.1018(5)   &	1.487(13)  & 1.3266(21) & 1.961(15)  & 1.11(2) \\
			$d$ (pc)												&  169.35(69) &   65.81(24)    &  87.25(57)  & 167.99(65)  &  76.64(37) & 92.21(71)\,\tablefootmark{a} \\
			$\pi$ (mas)												&   5.905(24) &    15.200(56)    &  11.462(74)  & 5.953(23)  &  13.049(63) & 10.845(83)\,\tablefootmark{a} \\
			\hline& & & 
		\end{tabular}
		\tablefoot{Values in parentheses are uncertainties on the final digits. $P_\mathrm{orb}$: orbital period. $T_\mathrm{p}$: time passage through periastron ($-2445000$). $e$: eccentricity. $K_1, K_2$: radial velocity semi-amplitude of the primary and secondary. $\gamma$: systemic velocity. $\omega$: argument of periastron. $\Omega$: position angle of the ascending node. $a$: semi-major axis. $i$: orbital inclination. $M_1, M_2$: mass of primary and secondary. $d, \pi$: distance and parallax.
			\tablefoottext{a}{From Gaia, taking into account the $0.029\,\mu$as offset.}
			\tablefoottext{b}{These two values correspond to a difference in systemic velocity, respectively for the primary and secondary (see Sect.~\ref{subsection__kwhya}).}
		}
		\label{table__orbits}
	\end{table*}
	
	We first fitted our astrometric measurements with radial velocities from \citet{Heminiak_2009_12_0} only. The r.m.s. of the velocity residuals is similar to \citet{Heminiak_2009_12_0}, we have 54 and 21\,\ms for the primary and secondary, respectively. The orbital elements are in good agreements with the latest published values \citep{Kirkby-Kent_2016_06_0,Sybilski_2018_07_0}, except for the inclination, for which we found the symmetric value with respect to $90^\circ$. All previous works made use of photometric measurements (instead of astrometry as in our case), but photometry does not allow to obtain the true inclination, and by convention the value $< 90^\circ$ is always adopted. The masses we obtained are in very good agreement with the published values (within $1\sigma$), with a precision level of $\sim 0.11$\,\%. The accuracy of the distance is also improved to $0.48$\,\%. We then performed the same fit with the velocities of \citet{Sybilski_2018_07_0} only. The r.m.s. of the velocity residuals is slightly larger with 88 and 40\,\ms for the primary and secondary, respectively. Except the systemic velocity, all derived parameters are in agreement (within $1\sigma$) with our previous estimates, including the inclination. As there are more velocity measurements, the precision is slightly better to $\sim0.09$\,\% for the masses, and 0.43\,\% for the distance. We only noticed a shift of 4.14\,\kms of the systemic velocity, but it does not impact the precision on the other parameters (which may comes from the orbital motion of the inner binary system (studied here) around the common centre of mass in this tertiary system). We note that the method used to derive the RVs by \citet{Heminiak_2009_12_0} and \citet{Sybilski_2018_07_0} are different, however, the masses are accurate with each other at $< 0.4\sigma$. \citet{Heminiak_2009_12_0} derived the RVs from the two-dimensional cross-correlation technique (TODCOR), while \citet{Sybilski_2018_07_0} used both TODCOR and the BF function, outside and during eclipses respectively. \citet{Sybilski_2018_07_0} also compared both methods and found the RVs to be in very good agreement. This demonstrates that the methods used to derive the RVs are robust and have not a significant impact on the measured masses.
	
	Our HARPS dataset spans a longer time range ($~ 7$\,yrs) than Helminiak and Sybilski's observations (6 and 1 month respectively), so we are more sensitive to the effect of the third component. The same fit is performed with our HARPS data and our astrometry. Clear offsets are detected for each data point and is related to the third component (rms$\sim 1.3\,$\kms). As our data set is limited, we manually searched for the best orbital period and eccentricity of a third component that gave the lowest residuals for the inner binary component. We found that a period around 109\,yrs with an eccentricity of $\sim 0.8$ provided the lowest residuals. We then corrected the velocities of the inner binary from the third component which we simultaneously fitted with our astrometry. The result is displayed in Fig.~\ref{figure__orbitaiphe}. We obtained a final r.m.s of 32\,\ms for the primary and 26\,\ms for the secondary. Our measured masses have a precision of 0.06\,\%, while the distance is precise at 0.41\,\%. We adopted these results as our final values, which are listed in Table~\ref{table__orbits}. We did not combined with the other datasets as the determination of the RVs was different and may lead to biases in the correction of the third component. However, it is worth mentioning that our masses are in very good agreement with our independent fit of Helminiak and Sybilski's data we previously performed with our astrometry, at $<0.7\sigma$ and $0.5\sigma$ respectively. This shows again the robustness of the methods in determining RVs.
	
	Our geometrical parallax is in agreement at $\sim 1.2\sigma$ with the parallax from GDR2. We note that we took into account the parallax offset of 0.029\,mas \citep{Lindegren_2018_08_0}. This slight disagreement might be explained by the fact that the GDR2 astrometric pipeline processing does not take into account yet the binarity of stars. Comparing with previous distance estimates ($162\pm6$\,pc from \citealt{Andersen_1988_05_0}, $168.4\pm6.8$\,pc from \citetalias{Graczyk_2017_03_0}), our value is within $1\sigma$, but we have a much better precision. We also note that the Hipparcos parallax of $3.50\pm1.04$\,mas is also in disagreement with our value, but the astrometric solutions also assumed a single star (probably reflected in the 30\,\% precision).

	\begin{figure*}[!h]
		\centering
		\resizebox{\hsize}{!}{\includegraphics{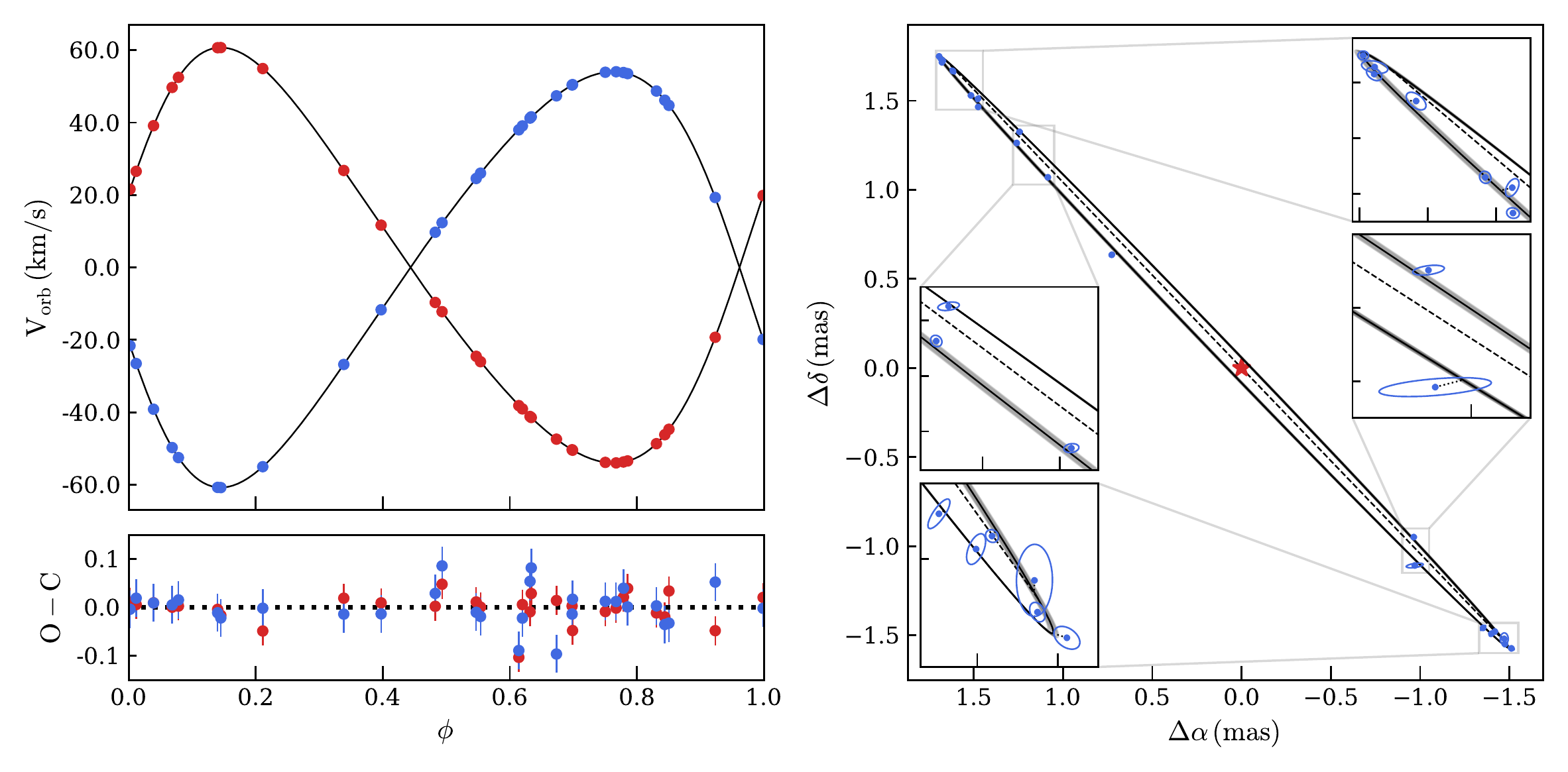}}
		\caption{\textit{Left}: radial velocities of the primary (red) and the secondary (blue) star in the AL~Dor system. \textit{Right}: PIONIER astrometric orbit. The shaded grey area represents the $1\sigma$ orbit.}
		\label{figure__orbitaldor}
	\end{figure*}
	\begin{figure*}[!h]
		\centering
		\resizebox{\hsize}{!}{\includegraphics{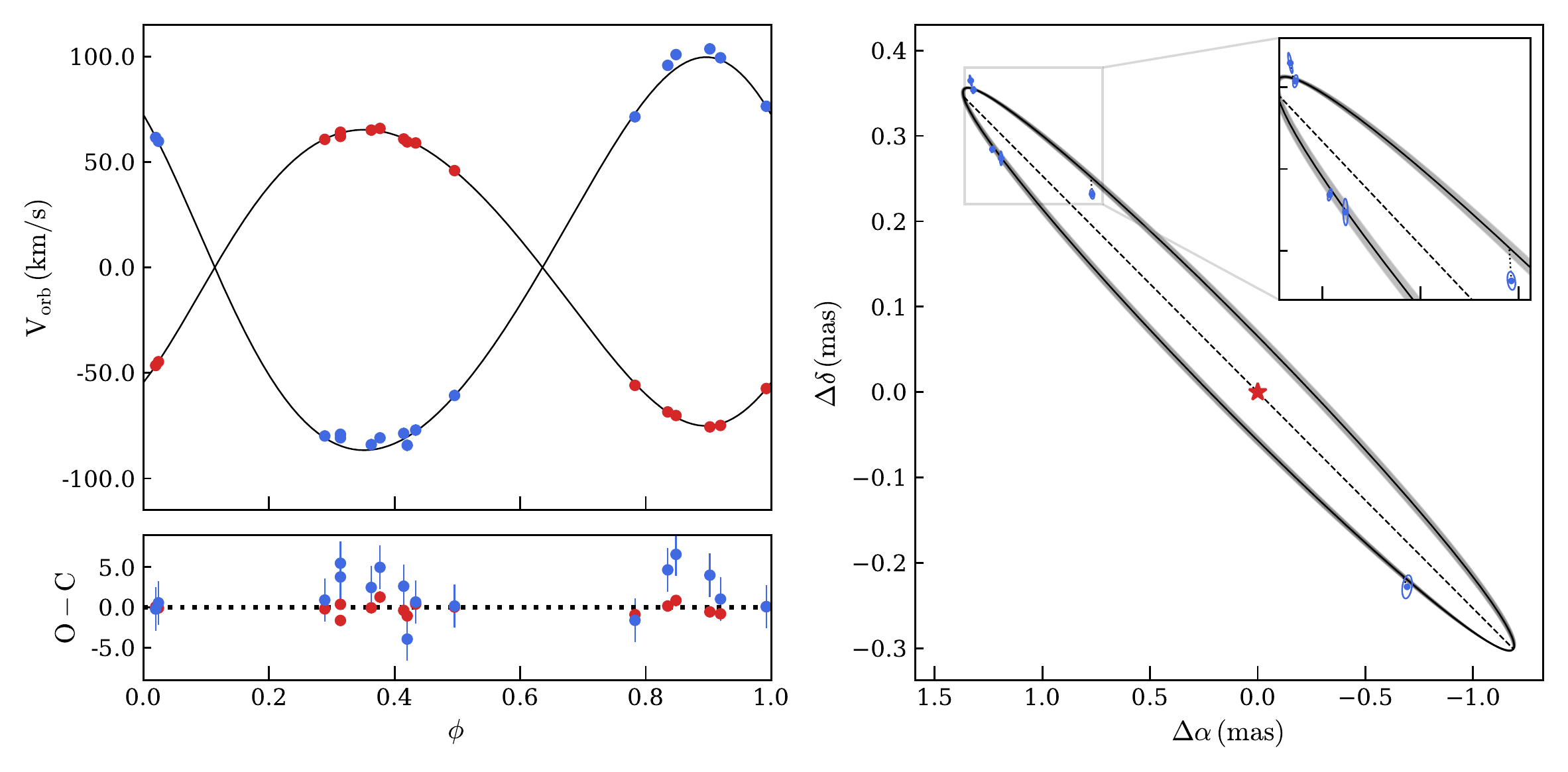}}
		\caption{\textit{Left}: radial velocities of the primary (red) and the secondary (blue) star in the KW~Hya system. \textit{Right}: PIONIER astrometric orbit. The shaded grey area represents the $1\sigma$ orbit.}
		\label{figure__orbitkwhya}
	\end{figure*}
	
	\subsection{AL~Doradus}
	
	We fitted our astrometric positions with RVs taken from \citet{Graczyk_2019_02_0}. We have an r.m.s. for the residual of the velocities of 30\,\ms and 39\,\ms for the primary and secondary, respectively. Our orbital fit is displayed in Fig.~\ref{figure__orbitaldor} and the resulting parameters are listed in Table~\ref{table__orbits}. For this system, the Gaia DR2 parallax is consistent with our value at a $0.6\sigma$ level (taking into account the $0.029$\,mas parallax offset). Although our precision on the distance is slightly lower than Gaia, our $0.37$\,\% precision is still competitive for checking the next Gaia data release as it includes all known systematics.
	
	We also precisely measured the mass of both components with a precision of 0.04\,\%. We found that both stars have the same mass, which is expected as the stars have similar spectral type.
	
	\subsection{KW~Hydrae}
	\label{subsection__kwhya}
	
	\begin{figure*}[!h]
		\centering
		\resizebox{\hsize}{!}{\includegraphics{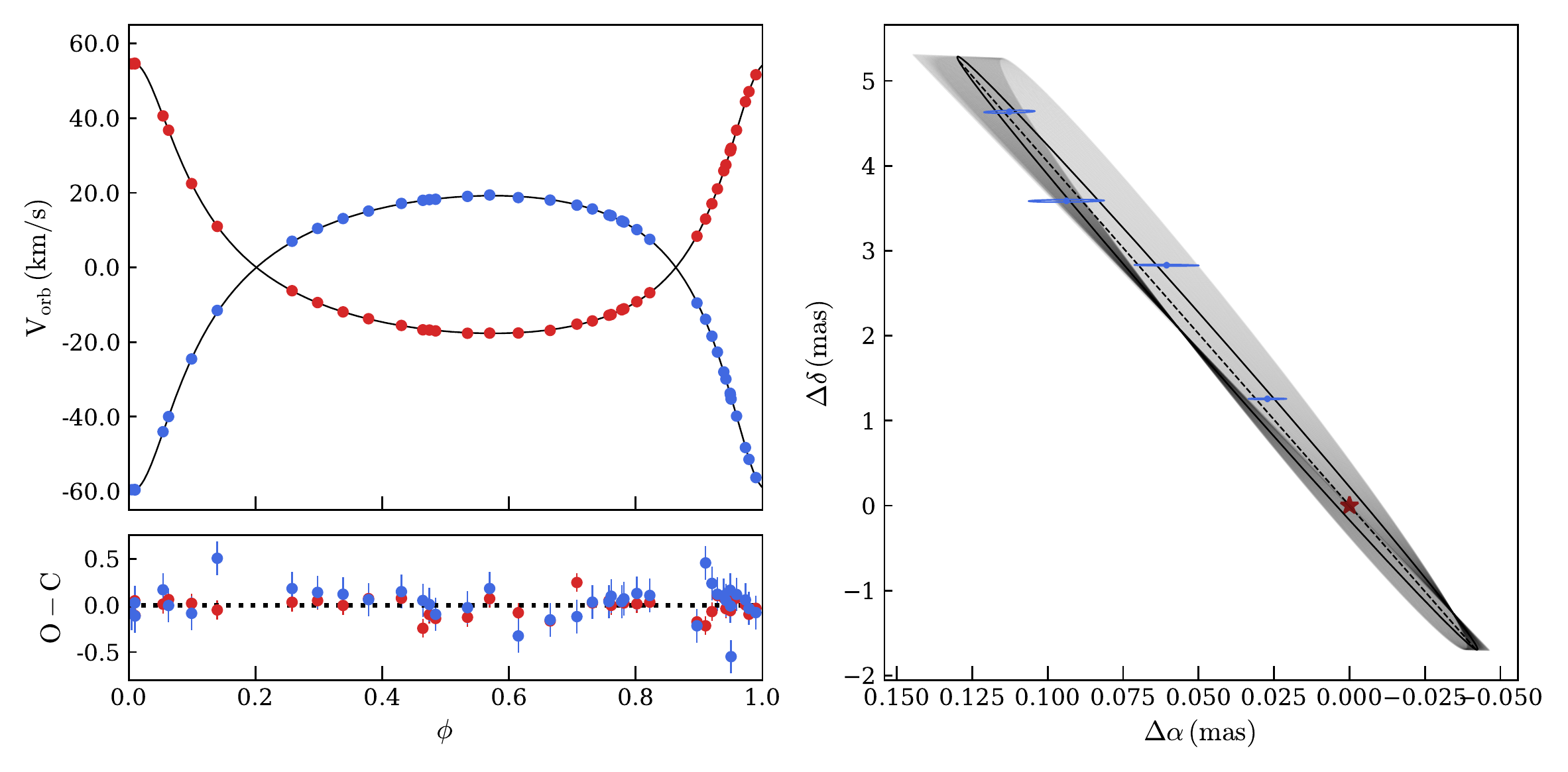}}
		\caption{\textit{Left}: radial velocities of the primary (red) and the secondary (blue) star in the NN~Del system. \textit{Right}: PIONIER astrometric orbit. The shaded grey area represents the $1\sigma$ orbit.}
		\label{figure__orbitnndel}
	\end{figure*}
	
	Our simultaneous fit is displayed in Fig.~\ref{figure__orbitkwhya} and our adopted parameters are listed in Table~\ref{table__orbits}. The r.m.s scatter in velocity of the secondary is not that good with 2.7\,\kms, while it is better for the primary with 0.7\,\kms, but similar to the values from \citet{Andersen_1984_01_0}. We noticed that the velocities of the secondary are shifted by about 2.5\,\kms, so for here we allowed to fit a separate systemic velocity. Such large shift is unlikely to be due to convective blueshift as observed sometimes \citep[see e.g.][]{Pilecki_2018_07_0,Gallenne_2016_02_0}, but probably due to the determination of the radial velocities. Our orbital parameters are consistent with the estimates of \citet{Andersen_1984_01_0}, except the orbital inclination where we found the solution $> 90^\circ$, for the same reason as we explained before for AI~Phe. Our derived masses are also in very good agreement, but we have a slightly better precision below 1\,\% for both stars. 
	
	We also obtained an orbital parallax with a precision of 0.7\,\%. The Gaia parallax is $\sim 2.1\sigma$ larger than our value (taking into account the parallax offset), probably because of the binarity as previously explained. However, the parallax from Hipparcos ($11.53\pm0.42$\,mas, assuming astrometric solutions for a single star) and \citet[][$11.63\pm0.54$\,mas]{Andersen_1984_01_0} are consistent within $1\sigma$ with our value.

	\subsection{NN~Delphini}
	
	We first fitted our astrometry with RVs of \citet{Griffin_2014_06_0}. We obtained a velocity r.m.s of 0.48\,\kms and 0.87\,\kms, for the primary and secondary respectively. We obtained masses precise to $\sim 0.9$\,\% and in very good agreement with the estimates of \citet{Griffin_2014_06_0}, within $0.4\sigma$. The estimated distance with these RVs only is $167.92\pm0.90$\,pc, in agreement with the Hipparcos value, but not with Gaia. We then fitted RVs of \citet{Sybilski_2018_07_0} only. We obtained a better velocity r.m.s of 17\,\ms and 130\,\ms, for the primary and secondary respectively (although there are only 7 measurements). All parameters are in agreement with our previous fit (within $1\sigma$, except the systemic velocity). The precision on the masses is $\sim0.9$\,\%, and consistent with Sybilski's values at $\sim 1.5\sigma$. The slight disagreement is because we rescaled the errorbars to the average scatter. The distance is also in agreement at a $0.3\sigma$ level with our previous estimate. We only noticed a velocity zero-point difference of $-1.235$\,\kms with the first dataset. As this system is not known to be tertiary, the shift may come from the use of a different instrument or a template with some intrinsic non-zero radial velocity. We also fitted our STELLA spectra alone. The velocity r.m.s obtained is 0.10\,\kms and 0.18\,\kms, for the primary and secondary respectively. Our fitted masses and distance are well within $1\sigma$ with the two previous analysis, except that the precision is much better. We derived masses precise to $\sim 0.18$\,\% and the distance to 0.4\,\%. Combining with Griffin's data degrades the precision because of the scatter, while the use with Sybilski measurements only slightly improve the precision (by 0.01\,\%), so we decided to use only our STELLA RVs. Our final fitted parameters are listed in Table~\ref{table__orbits}, and the orbit is displayed in Fig.~\ref{figure__orbitnndel}. We derived the mass of both components precise to $\sim 0.2$\,\%. Our derived final parallax is precise to 0.4\,\%, and is $4.2\sigma$ larger than the Gaia parallax (parallax offset included), but in agreement with the Hipparcos estimate (measured assuming a single star).
	
	\subsection{$\psi$~Centauri}
	
\begin{figure*}[!h]
		\centering
		\resizebox{\hsize}{!}{\includegraphics{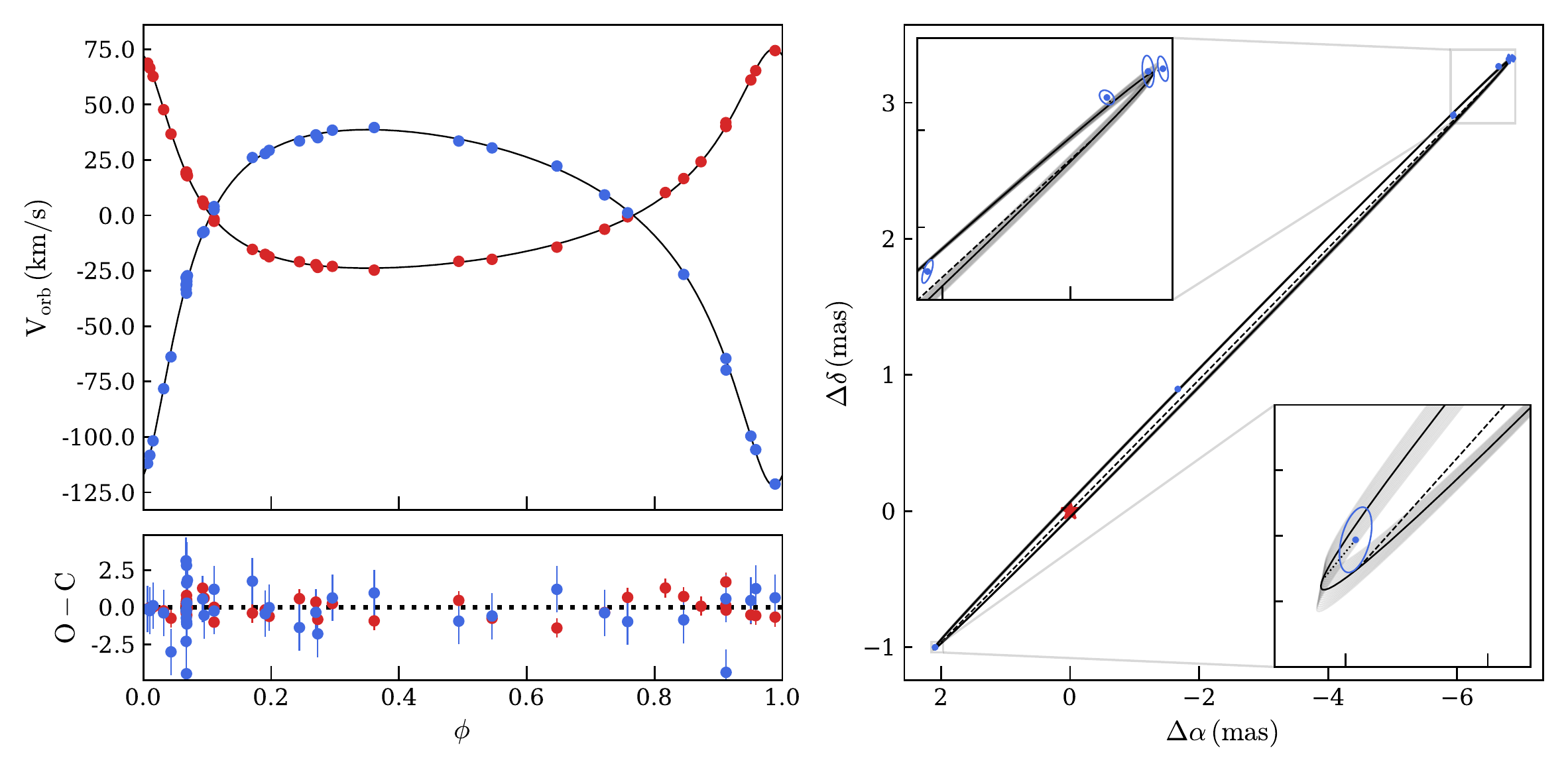}}
		\caption{\textit{Left}: radial velocities of the primary (red) and the secondary (blue) star in the $\psi$~Cen system. \textit{Right}: PIONIER astrometric orbit. The shaded grey area represents the $1\sigma$ orbit.}
		\label{figure__orbitpsicen}
	\end{figure*}

	We displayed our combined fit in Fig.~\ref{figure__orbitpsicen} and our final adopted parameters in Table~\ref{table__orbits}. We obtained an r.m.s. on the velocities of 0.64\,\kms for the primary and 1.57\,\kms for the secondary. We measured precise masses at a $\sim 0.9$\,\% level, which agrees with the estimates of \citep{Bruntt_2006_09_0}. However, they are at $\sim 2-3\sigma$ higher than the estimates of \citet{Mantegazza_2010_03_0}. This difference probably comes from our different and more adapted way of extracting the radial velocities from the spectra. This is noticed in our estimates of $K_1$ and $K_2$ which are different by $2.6\sigma$ and $1.6\sigma$, respectively.
	
	We derived a distance accurate to 0.5\,\%, which is larger by $2.1\sigma$ with the Hipparcos value. This is possibly due to the astrometric solutions of Hipparcos which assumed a single star. The Gaia parallax is $3.1\sigma$ larger than our value (parallax offset included), and probably also because the pipeline assumed a single star for this system.

	\subsection{V4090~Sagittarii}
	
\begin{figure*}[!h]
		\centering
		\resizebox{\hsize}{!}{\includegraphics{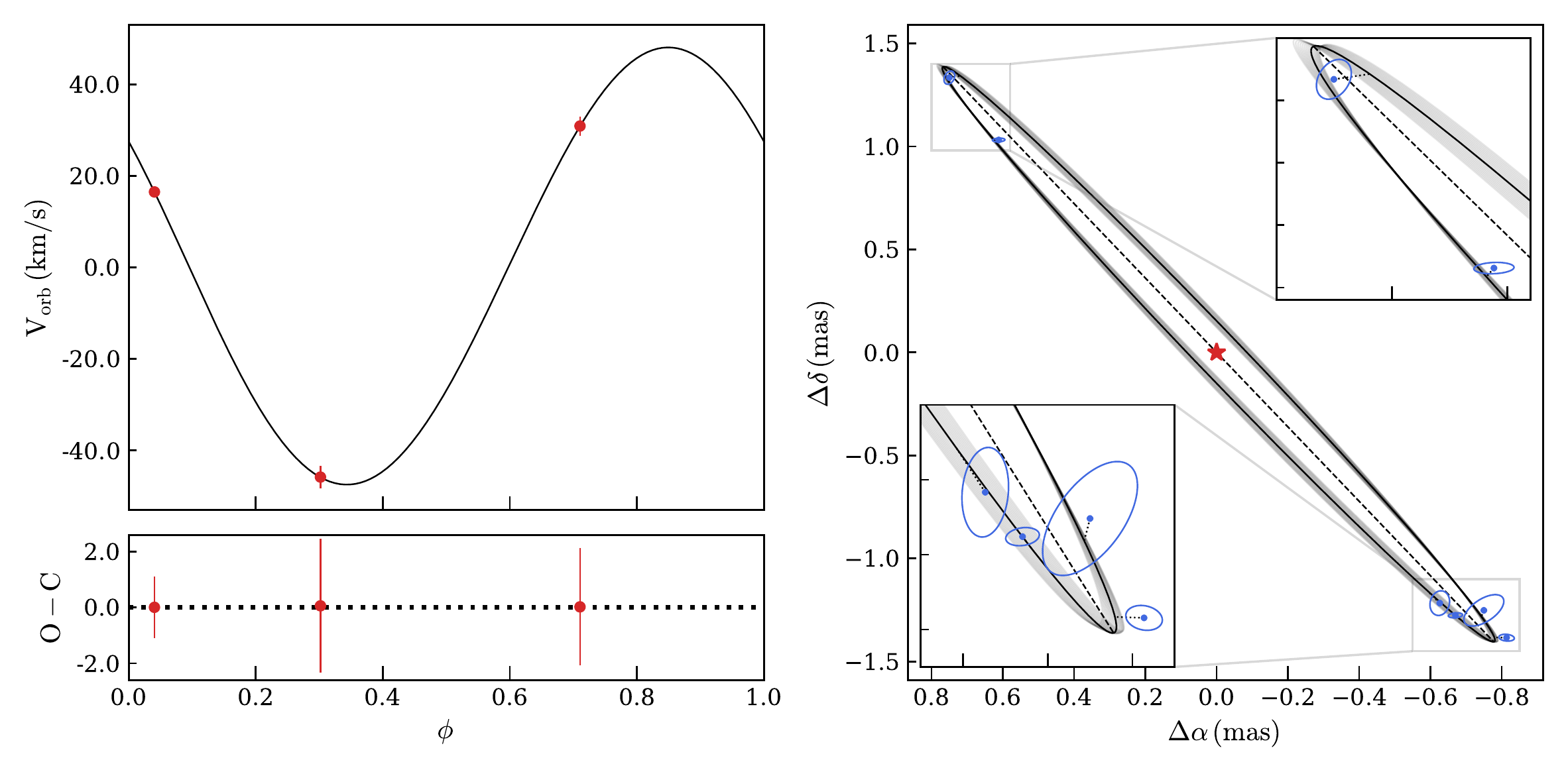}}
		\caption{\textit{Left}: radial velocities of the primary in the V4090~Sgr system. \textit{Right}: PIONIER astrometric orbit. The shaded grey area represents the $1\sigma$ orbit.}
		\label{figure__orbitv4090sgr}
	\end{figure*}

	As this system is still a single-line spectroscopic binary, our astrometric positions were fitted with the radial velocities of the primary star only \citep{Nordstrom_1985_07_0}. Our final fit is displayed in Fig.~\ref{figure__orbitv4090sgr} and the resulting parameters are listed in Table~\ref{table__orbits}. To estimate the masses from a single line binary, we must assume the distance to the system. We adopted the Gaia distance as it is at $1\sigma$ with the Hipparcos value. In our MCMC procedure, we included the parallax uncertainty using a normal distribution centred on 10.8450\,mas with a standard deviation of 0.083\,mas. Our derived masses are reported in Table~\ref{table__orbits}, and have a precision of $\sim 3\,$\%. They are in agreement (within $2\sigma$) with the estimates of \citet{North_1997_08_0}, but here we measured the dynamical masses instead of inferring it from evolution models. 
	
	\section{Discussion}
	\label{section__discussion}

	\subsection{Mass comparison with the literature}
	
\begin{figure*}[!h]
		\centering
		\resizebox{\hsize}{!}{\includegraphics[width = \linewidth]{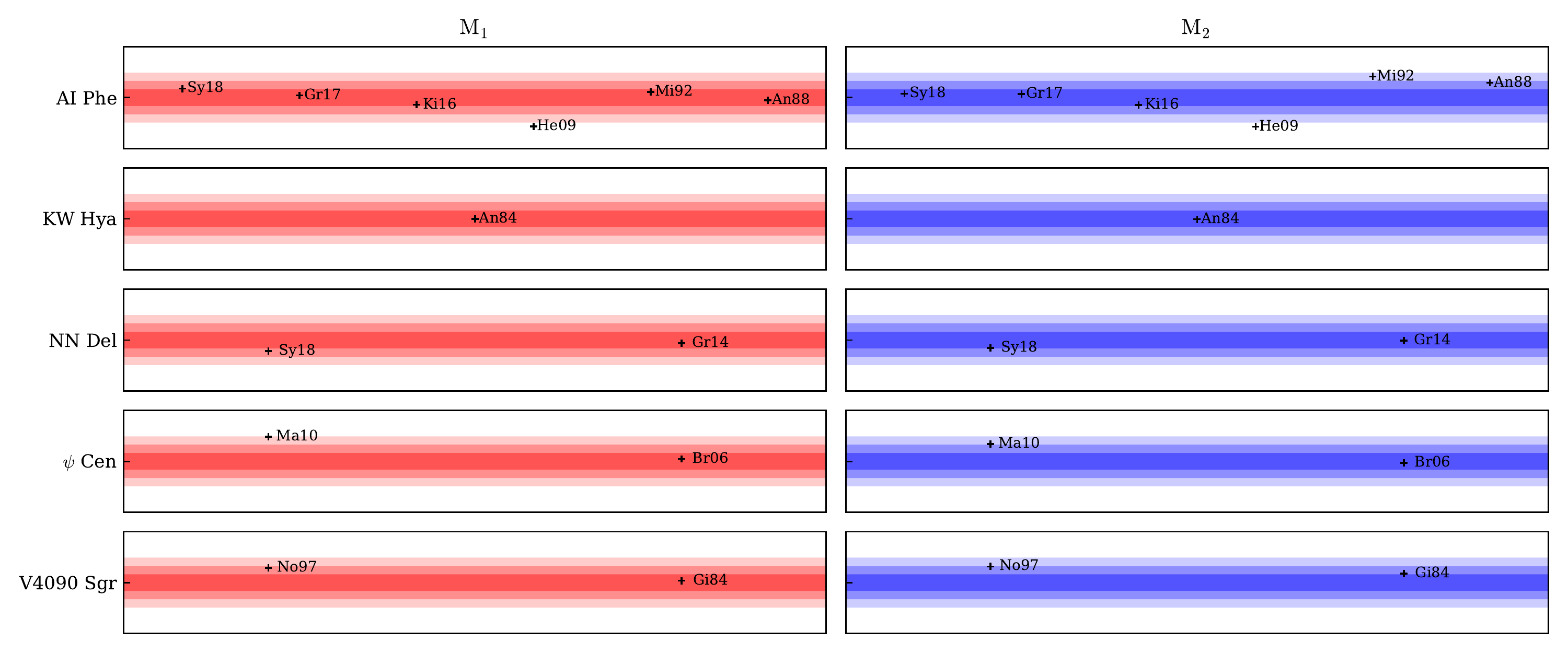}}
		\caption{Comparison of our mass measurements (coloured area) with the literature (black dots). The red (left) and blue (right) shaded areas represent the $1\sigma, 2\sigma$ and $3\sigma$ combined uncertainties of the primary and secondary, respectively. References are: Sy18 for \citet{Sybilski_2018_07_0}, Gr17 for \citet{Graczyk_2017_03_0}, Ki16 for \citet{Kirkby-Kent_2016_06_0}, He09 for \citet{Heminiak_2009_12_0}, Mi92 for  \citet{Milone_1992_03_0}, An88 for \citet{Andersen_1988_05_0}, An84 for  \citet{Andersen_1984_01_0}, Gr14 for \citet{Griffin_2014_06_0},  Ma10 for \citet{Mantegazza_2010_03_0}, Br06 for \citet{Bruntt_2006_09_0}, No97 for \citet{North_1997_08_0} and Gi84 for \citet{Giuricin_1984_06_0}.}
		\label{figure_mass_comparison}
	\end{figure*}

	In Fig.~\ref{figure_mass_comparison} we present a comparison of our mass measurements with previous estimates from the literature, except for AL~Dor for which there are no existing measurements. We note that all literature data are based on the combination of RVs and photometric light curves which are more dependent of atmospheric models.
	
	We see that most previous measurements are within $1-2\sigma$ with ours, but we note a few discrepancies $> 2\sigma$. The masses of the AI~Phe components estimated by \citet{Heminiak_2009_12_0} are more than $3\sigma$ away. In Sect.~\ref{subsection__results_aiphe}, we fitted our astrometry with only their RVs and we found masses in agreements with our final values at $<0.7\sigma$. We therefore conclude that the discrepancy might come from the fit of the photometric light curves and the fact that they only have eight velocity measurements. The masses estimated by \citet{Mantegazza_2010_03_0} for the $\psi$~Cen system are also $> 2\sigma$ with our values and the estimate from \citet[][estimated from isochrone fitting]{Bruntt_2006_09_0}. As explained previously in Sect.~\ref{section__observations}, in the method used by \citet{Mantegazza_2010_03_0}, individual RVs are not calculated, and as they stated, they did not completely disentangle the components, which probably add some errors. We also note that they did not perform a combined fit with photometric light curves.
	
	\subsection{Evolutionary state}
	
	We employed the same fitting method as in \citet{Gallenne_2018_08_0,Gallenne_2016_02_0}, i.e. we fitted the \parsec \citep[PAdova and TRieste Stellar Evolution Code,][]{Bressan_2012_11_0}, \basti \citep[Bag of Stellar Tracks and Isochrones,][]{Pietrinferni_2004_09_0} and \mist \citep[Mesa Isochrones and Stellar Tracks,][]{Choi_2016_06_0} isochrone models to estimate the stellar age of our systems. These models are well suited as they include the horizontal and asymptotic giant branch evolutionary phases, and contain a wide range of initial masses and metallicities. In addition, it enable us to test the uncertainty of the age induced by different stellar models.
	
	\parsec models are computed for a scaled-solar composition with $Z_\odot=0.0152$, and follow a helium initial content relation $Y = 0.2485 + 1.78Z$ with a mixing length parameter $\alpha_\mathrm{MLT} = 1.74$. They include convective core overshooting during the main sequence phase, parametrized with the strength of convective overshooting in units of the pressure scale height $l_\mathrm{ov} = \alpha_\mathrm{ov}H_\mathrm{p}$. The overshooting parameter $\alpha_\mathrm{ov}$ is set depending on the mass of the star, i.e.  $\alpha_\mathrm{ov} = 0$ for $M \lesssim 1.1\,M_\odot$, $\alpha_\mathrm{ov} \sim 0.25$ for $M \gtrsim 1.4\,M_\odot$, and linearly ramps with the mass in between. The \basti models are computed for a scaled-solar composition with $Z_\odot=0.0198$, following the relation $Y = 0.245 + 1.4Z$ with $\alpha_\mathrm{MLT} = 1.913$. They also include convective core overshooting with the same parametrization, but with the conditions $\alpha_\mathrm{ov} = 0$ for $M \lesssim 1.1\,M_\odot$, $\alpha_\mathrm{ov} = 0.20$ for $M \gtrsim 1.7\,M_\odot$, and $(M - 0.9M_\odot)/4$ in between. The \mist models used a scaled-solar composition with $Z_\odot=0.0142$, with the relation $Y = 0.2703 + 1.5Z$ and $\alpha_\mathrm{MLT} = 1.82$. They used an alternate prescription of the core overshooting with a diffusion coefficient $D_\mathrm{ov} = D_0\exp{(-2z/H_\nu)}$, where $z$ is the distance from the edge of the convective zone, $D_0$ the coefficient at $z = 0$, and $H_\nu$ is defined with the overshooting parameter $f_\mathrm{ov}$ such that $H_\nu = f_\mathrm{ov}H_\mathrm{p}$. \mist models adopted a fixed value $f_\mathrm{ov} = 0.016$ for all stellar masses, which would be approximatively converted to $\alpha_\mathrm{ov} \sim 0.18$ \citep{Claret_2017_11_0}.
	
	
	We retrieved several isochrones from the \parsec database tool\footnote{\url{http://stev.oapd.inaf.it/cgi-bin/cmd}}, with ages ranging from $\log t = 6.6$ to 10 by step of 0.05 (i.e. $\sim 0.1$-13\,Gyr), and metallicities from $Z = 0.003$ to 0.06 (i.e. $-0.7 \leq \mathrm{[Fe/H]} \leq +0.6$, using $[\mathrm{Fe/H}] \sim \log{(Z/Z_\odot)}$), by step of 0.001 (fine enough to avoid re-interpolation). The \basti isochrones are pre-computed in their database\footnote{\url{http://basti.oa-teramo.inaf.it/index.html}}, we downloaded models for $t = 0.1-9.5$\,Gyr by step of $\sim 0.2$\,Myr and $Z = 0.002, 0.004, 0.008, 0.01, 0.0198, 0.03$ and 0.04 (i.e. $-1.0 \leq \mathrm{Fe/H} \leq 0.3$). For fitting purpose, we created an interpolated grid of the \basti isochrones in $Z$, from 0.002 to 0.04 by step of 0.001. We also computed \mist isochrones from their database tool\footnote{\url{http://waps.cfa.harvard.edu/MIST/interp_isos.html}} using the standard age grid from 0.1\,Myr to 20\,Gyr by step of $\sim 1$\,Myr, and for metallicities in the range $0.002 \leq Z \leq 0.04$ (i.e. $-0.85 \leq \mathrm{Fe/H} \leq 0.45$) by step of 0.001.
	
	When possible, we searched for the best-fit age in stellar effective temperature, radius and mass for both component simultaneously, assuming coeval stars and following a $\chi^2$ statistics
	\begin{displaymath}
	\chi^2 = \sum_{i=1}^2 \left[ \left(\dfrac{\Delta T_\mathrm{eff}}{\sigma_\mathrm{T_{eff}}}\right)_i^2 +
	\left(\dfrac{\Delta R}{\sigma_R}\right)_i^2 + \left(\dfrac{\Delta M}{\sigma_M}\right)_i^2  \right]  
	\end{displaymath}
	where the sum is over both components ($i = 1, 2$) and the $\Delta$ symbol represents the difference between the predicted and observed quantities. The effective temperature and the radii are measured quantities and were taken from the literature. They are listed in Table~\ref{table__atmospheric_parameter}.  The masses are also measured from this work and are reported in Table~\ref{table__orbits}. We took care of rescaling the retrieved linear radii according to our own estimate of the linear semi-major axis. In our isochrone plots we also displayed the stellar luminosity estimated from the Stefan-Boltzmann law, but this parameter was not including in the fit as this is not an independent measurements.
	
	The stellar metallicity is kept fixed in this process to a value from the literature (listed in Table~\ref{table__atmospheric_parameter}). Our fitting procedure was the following.
	For all isochrone models, we first chose the closest grid in $Z$ for a given metallicity. Then, we searched for the global $\chi^2$ minimum in age by fitting all isochrones for that given metallicity. A second fit is then performed around that global minimum value, and where the grid is interpolated in age at each iteration. To assess the uncertainties on the three isochrone models (i.e. \parsec, \basti and \mist), we repeated the process with $Z \pm \sigma$. Our final adopted age corresponds to the average and standard deviation between the three models.
	
	
	\paragraph{AI~Phoenicis}
	
	\begin{figure}[!h]
		\centering
		\resizebox{\hsize}{!}{\includegraphics[width = \linewidth]{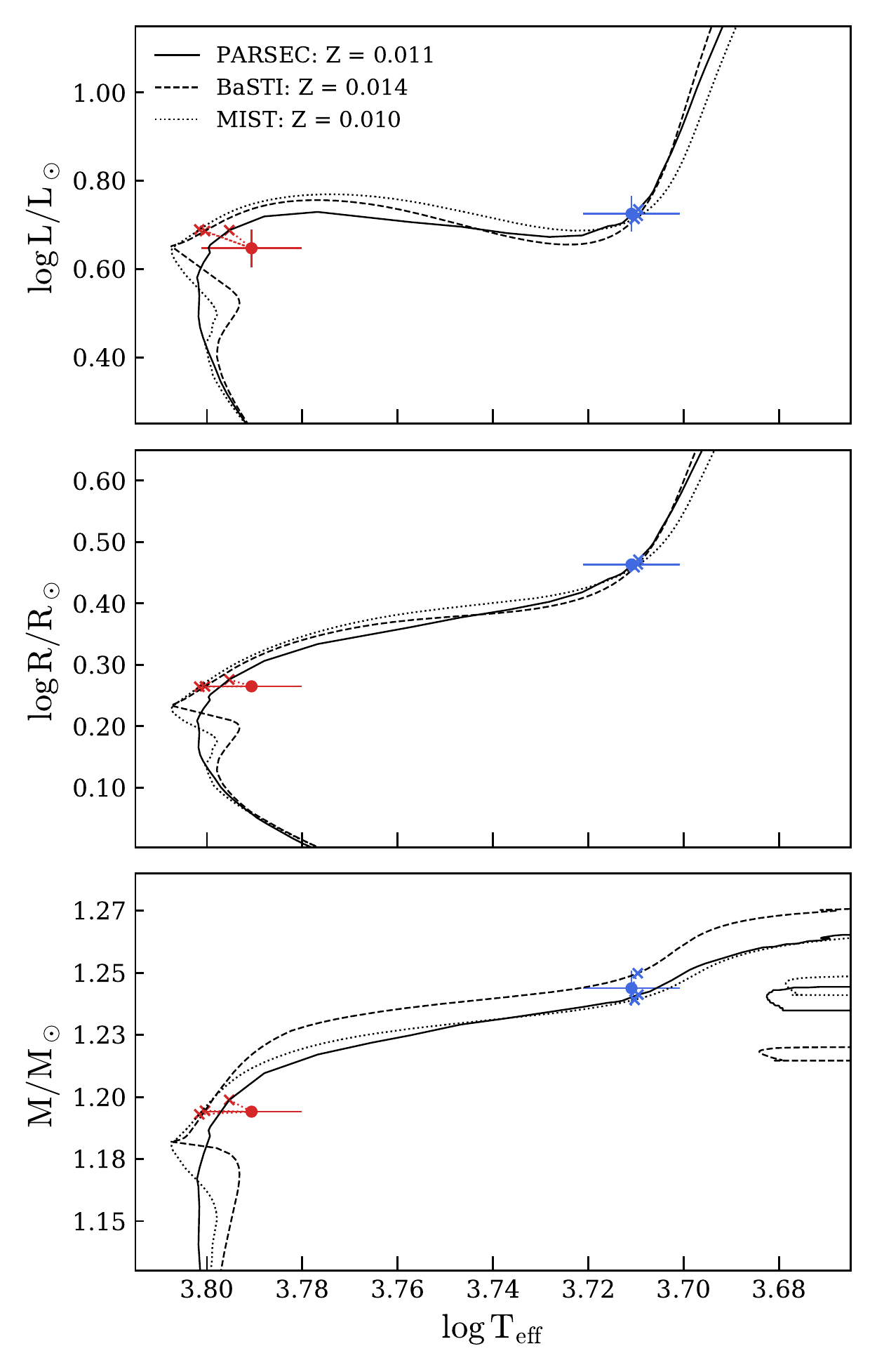}}
		\caption{Fitted \parsec, \basti and \mist isochrones for the AI~Phe system. We note that the luminosity is not fitted and is estimated from the Stefan-Boltzmann law.}
		\label{figure_isochrones_aiphe}
	\end{figure}
	
	We adopted the metallicity from \citet{Andersen_1988_05_0} and the stellar parameters listed in Table~\ref{table__atmospheric_parameter}. All fitted isochrones are consistent and give a similar age within $1\sigma$ (see Table~\ref{table__atmospheric_parameter}). They are displayed in Fig.~\ref{figure_isochrones_aiphe}. We found an average age of this system of $t = 4.20\pm0.19$\,Gyr. This value is in good agreement with previous estimates of \citet[$4.39\pm0.32$\,Gyr,][]{Kirkby-Kent_2016_06_0}, \citet[4.3\,Gyr,][]{Kanjanascul_2012_06_0}, \citet[4\,Gyr,][]{Heminiak_2009_12_0} \citet[$4.1\pm0.4$\,Gyr,][]{Andersen_1988_05_0}, who used different stellar evolution models. The more evolved component is particularly in very good agreement for all models, while the other component located at the turn-off point disagrees at $> 3\sigma$ with the \parsec model. This might be linked to differences in the implementation of overshooting, Helium content or the mixing length in the models. We note however that the strength of convective core overshooting starts increasing rapidly from $\sim 1.2\,M_\odot$ \citep{Claret_2016_07_0,Claret_2017_11_0,Claret_2018_06_0}, so for this system with masses $< 1.25\,M_\odot$ the effect of overshooting should be negligible.
	
	
	
	\paragraph{AL~Doradus}
	
	\begin{figure}[!h]
		\centering
		\resizebox{\hsize}{!}{\includegraphics[width = \linewidth]{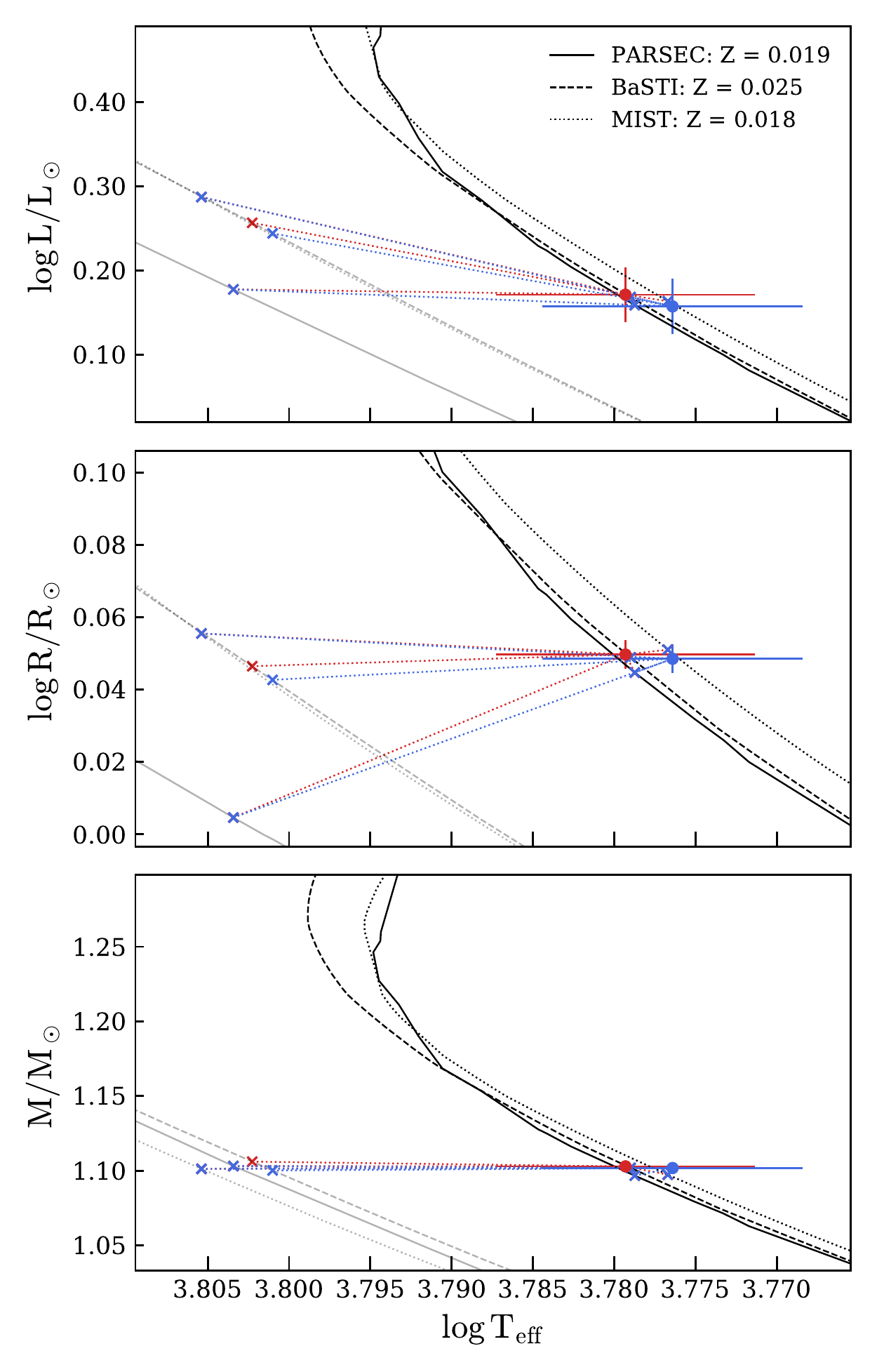}}
		\caption{Fitted \parsec, \basti and \mist isochrones for the AL~Dor system. The grey isochrones correspond to a metallicity of $\mathrm{[Fe/H]}= -0.21$\,dex, while the black isochrones are for 0.10\,dex. We note that the luminosity is not fitted and is estimated from the Stefan-Boltzmann law.}
		\label{figure_isochrones_aldor}
	\end{figure}
	
	We first used the metallicity $\mathrm{[Fe/H]}= -0.21$\,dex derived by \citet{Graczyk_2019_02_0} with the stellar parameters listed in Table~\ref{table__atmospheric_parameter}. We found that both stars are in a similar main-sequence phase, however, the best-fit isochrones are not satisfactory for all models, as shown in Fig.~\ref{figure_isochrones_aldor} in grey colour. The corresponding average age is $t = 1.28\pm0.89$\,Gyr. The metallicity does not seem consistent, and the value $\sim 0.10$\,dex would be more appropriate to reconcile the observables with the isochrones, as plotted in Fig.~\ref{figure_isochrones_aldor} in black colour. This value is consistent with the $0.15\pm0.15$\,dex derived by \citet{Ammons_2006_02_0} from a metallicity-colour calibration. All isochrone models locate the stars near the main-sequence turn-off with an average age $t_\mathrm{avg} = 2.97\pm0.23$\,Gyr. This can be compared to \citet{Casagrande_2011_06_0} who performed a Bayesian analysis of the primary star with the \parsec and  \basti isochrones to derive $t = 4.30\pm0.85$\,Gyr and $t = 3.90\pm0.90$\,Gyr, respectively. Although they used a $\sim100$\,K hotter temperature and a metallicity of $-0.10$\,dex, their estimates are consistent with ours at $< 1.5\sigma$.
	
	Radius are in agreement at $\sim1\sigma$ with all models, while the masses disagree at more than $2.5\sigma$, the \parsec model showing the largest discrepancy. A better estimate of the metallicity and more precise temperatures are needed for further conclusions.
	
	\paragraph{KW~Hydrae}
	
	\begin{figure}[!h]
		\centering
		\resizebox{\hsize}{!}{\includegraphics[width = \linewidth]{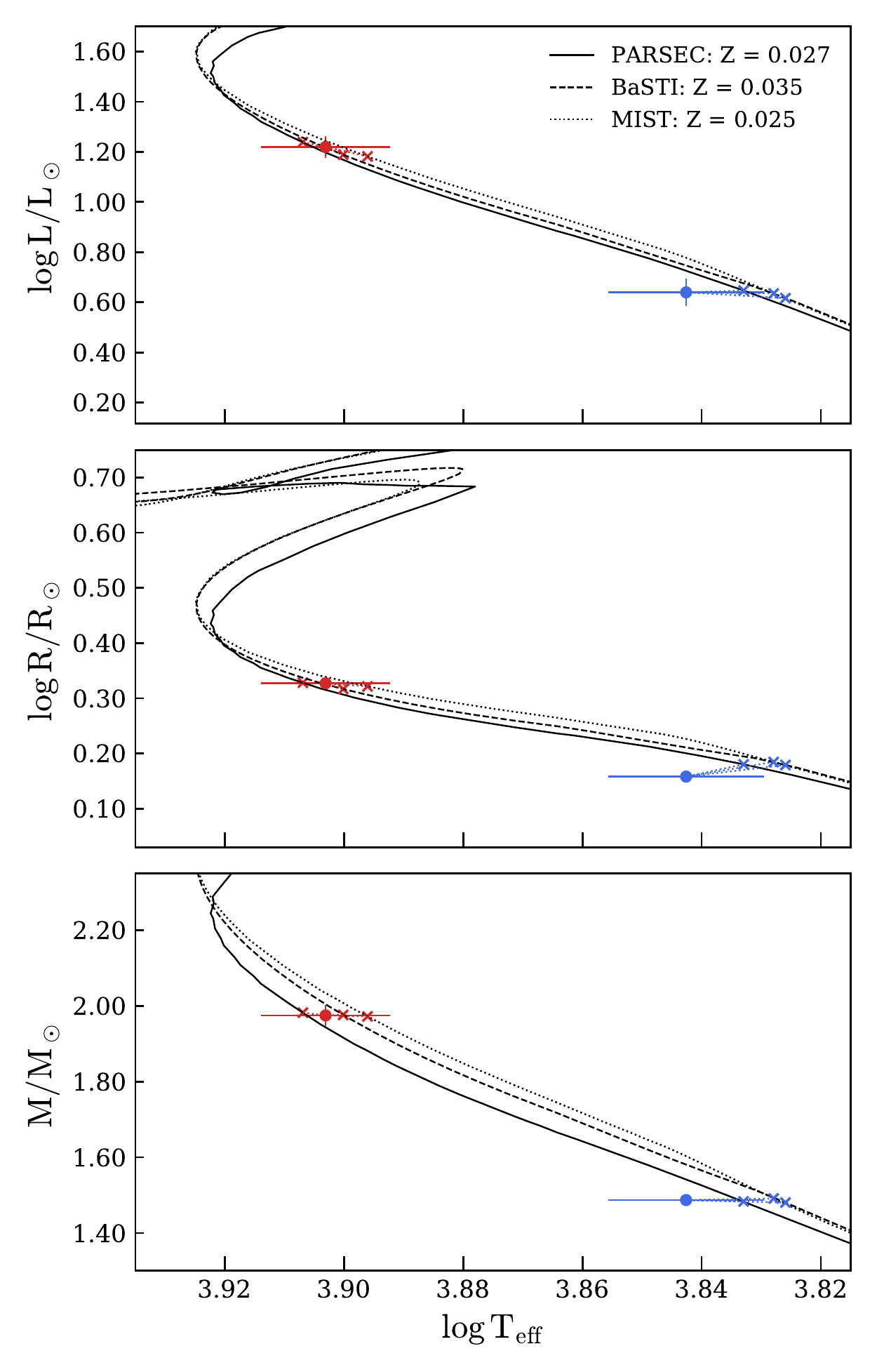}}
		\caption{Fitted \parsec, \basti and \mist isochrones for the KW~Hya system. We note that the luminosity is not fitted and is estimated from the Stefan-Boltzmann law.}
		\label{figure_isochrones_kwhya}
	\end{figure}
	
	There is no measurement of metallicity for this system in the literature. So we manually searched for a [Fe/H] value giving the lowest $\chi^2$ for all models. We found that $\mathrm{[Fe/H]}= 0.25$\,dex provided a satisfactory fit for all models with the stellar parameters listed in Table~\ref{table__atmospheric_parameter}. We conservatively chose an uncertainty of $\pm 0.10$\,dex to derive our age uncertainty. The isochrones are displayed in Fig.~\ref{figure_isochrones_kwhya}, and are all consistent with each other. They show two components located on the main sequence, with the more evolved one being near the turn-off point. The age of all model agrees and we adopted an average age for the KW~Hya system of $t_\mathrm{avg} = 0.54\pm0.05$\,Gyr. This is consistent with the 0.5\,Gyr derived by \citet{Andersen_1984_01_0} with $Z = 0.02$ and other isochrones models using old opacity data.
	
	However, we notice that the more evolved component (the primary) is better fitted than the less evolved star (the secondary). More particularly, the fit of the radius of the secondary is not fully satisfactory. Removing the mass measurements from the fitted parameters slightly improved the agreement in radius, which might mean that the $M-R$ calibration of the models is still not optimal for this level of precision. Finally, it is worth mentioning that fitting each component independently provides a similar age for the primary (within $1\sigma$), while the secondary is $\sim 7\sigma$ younger.
	
	\paragraph{NN~Delphini}
	
	\begin{figure}[!h]
		\centering
		\resizebox{\hsize}{!}{\includegraphics[width = \linewidth]{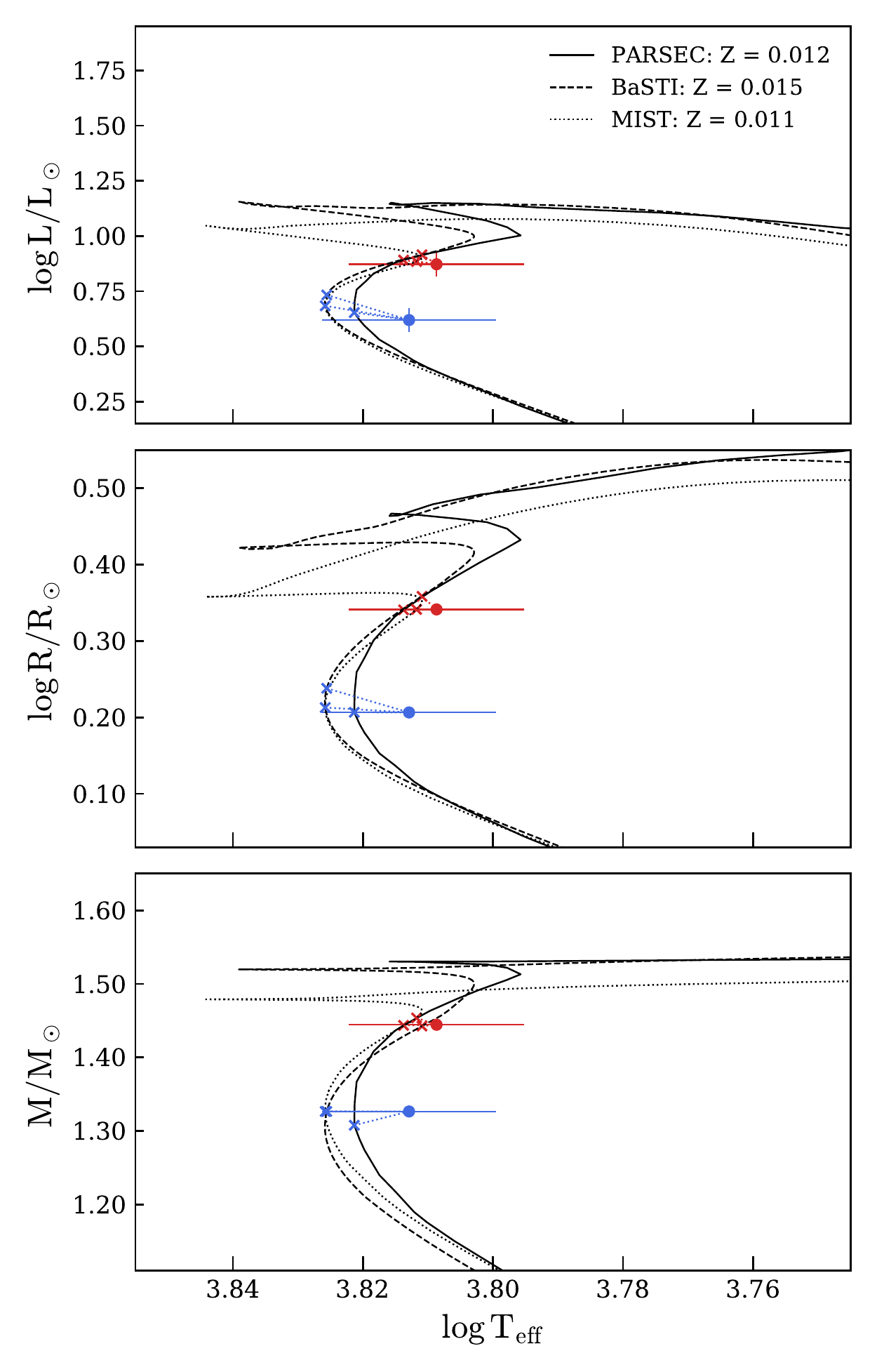}}
		\caption{Fitted \parsec, \basti and \mist isochrones for the NN~Del system. We note that the luminosity is not fitted and is estimated from the Stefan-Boltzmann law.}
		\label{figure_isochrones_nndel}
	\end{figure}
	
	There is no estimate of the metallicity for this system. So we did as for KW~Hya, i.e. we manually searched for a value giving the lowest $\chi^2$ for all models, given the parameters listed in Table~\ref{table__atmospheric_parameter}. We found that $\mathrm{[Fe/H]}= -0.10$\,dex provided a satisfactory fit for all models. We also conservatively chose an uncertainty of $\pm 0.10$\,dex to derive our age uncertainty for each models. The isochrones are displayed in Fig.~\ref{figure_isochrones_nndel}, and are in agreement with each other. All isochrones give the same evolutionary phase, i.e. both components are at the main-sequence turn-off point. The age between the \parsec and \mist models are very similar, while the \basti model gives a younger system. They are reported in Table~\ref{table__atmospheric_parameter}. We adopted the average $t_\mathrm{avg} = 2.21\pm0.18$\,Gyr.
	
	We also notice that our secondary mass measurement is $\sim9\sigma$ away from the \basti isochrones. Again, this might be linked to the calibrations of the models, but further analysis are still necessary, particularly a measurement of the metallicity.
	
	\paragraph{$\psi$~Centauri}
	
	\begin{figure}[!h]
		\centering
		\resizebox{\hsize}{!}{\includegraphics[width = \linewidth]{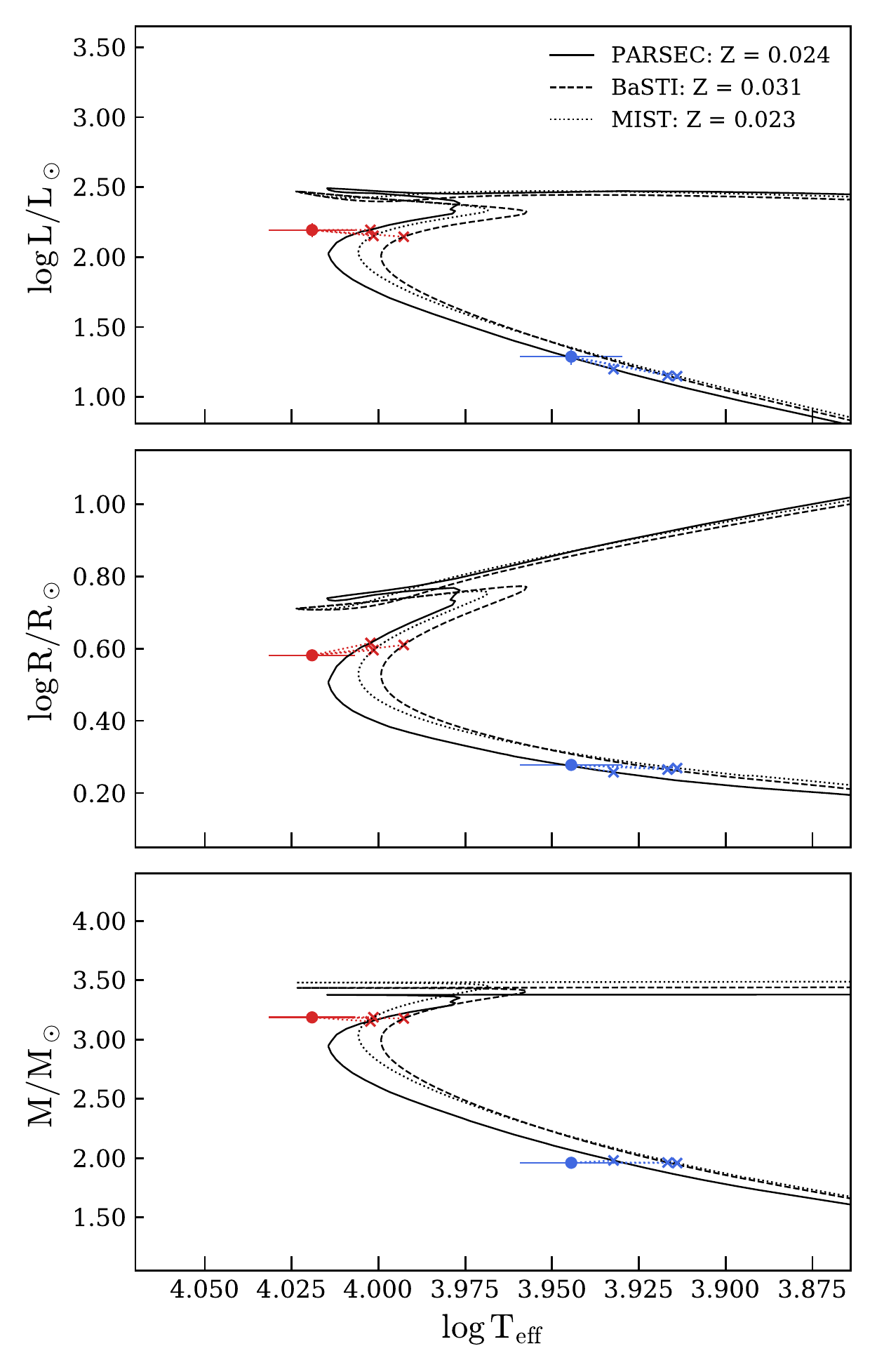}}
		\caption{Fitted \parsec, \basti and \mist isochrones for the $\psi$~Cen system. The grey isochrones correspond to a metallicity of $\mathrm{[Fe/H]}= 0.1$\,dex, while the black isochrones are for 0.05\,dex. We note that the luminosity is not fitted and is estimated from the Stefan-Boltzmann law.}
		\label{figure_isochrones_psicen}
	\end{figure}
	
	The metallicity of $\psi$~Cen is also unknown, so we did as our previous systems and manually searched for a value giving the lowest $\chi^2$ for all models, according to the parameters listed in Table~\ref{table__atmospheric_parameter}. We found that [Fe/H] = 0.2\,dex provided a satisfactory fit for all models. We also conservatively chose an uncertainty of $\pm 0.10$ dex to derive our age uncertainty. The isochrones are displayed in Fig.~\ref{figure_isochrones_psicen}. We found that the secondary is on the main-sequence while the primary entered the main-sequence turn-off. We estimated an average age of $0.28\pm0.01$\,Gyr, consistent with the estimate of \citet[][290\,Myr]{Bruntt_2006_09_0} who used Yonsei-Yale model isochrones and a metallicity $Z = 0.01812$. While all models give a similar age, the \parsec model gives a better fit with the measurements. We notice that the less evolved component is in better agreement with the tracks than the primary star located at the turn-off point. Changing the metallicity value does not reconcile the tracks, and the slight disagreement may be due to internal physics used in the models. The difference in the parametrization of the core overshooting may be a lead to explore. For instance, we notice that the \parsec model which has the largest overshooting parameter ($\alpha_\mathrm{ov} = 0.25$) provides a slightly better agreement than the other models. However, the \mist model with the lowest parameter ($\alpha_\mathrm{ov} \sim 0.18$) is in between, but they also used a alternate prescription of the core overshooting.
	
	\paragraph{V4090~Sagittarii}
	
\begin{figure}[!h]
		\centering
		\resizebox{\hsize}{!}{\includegraphics[width = \linewidth]{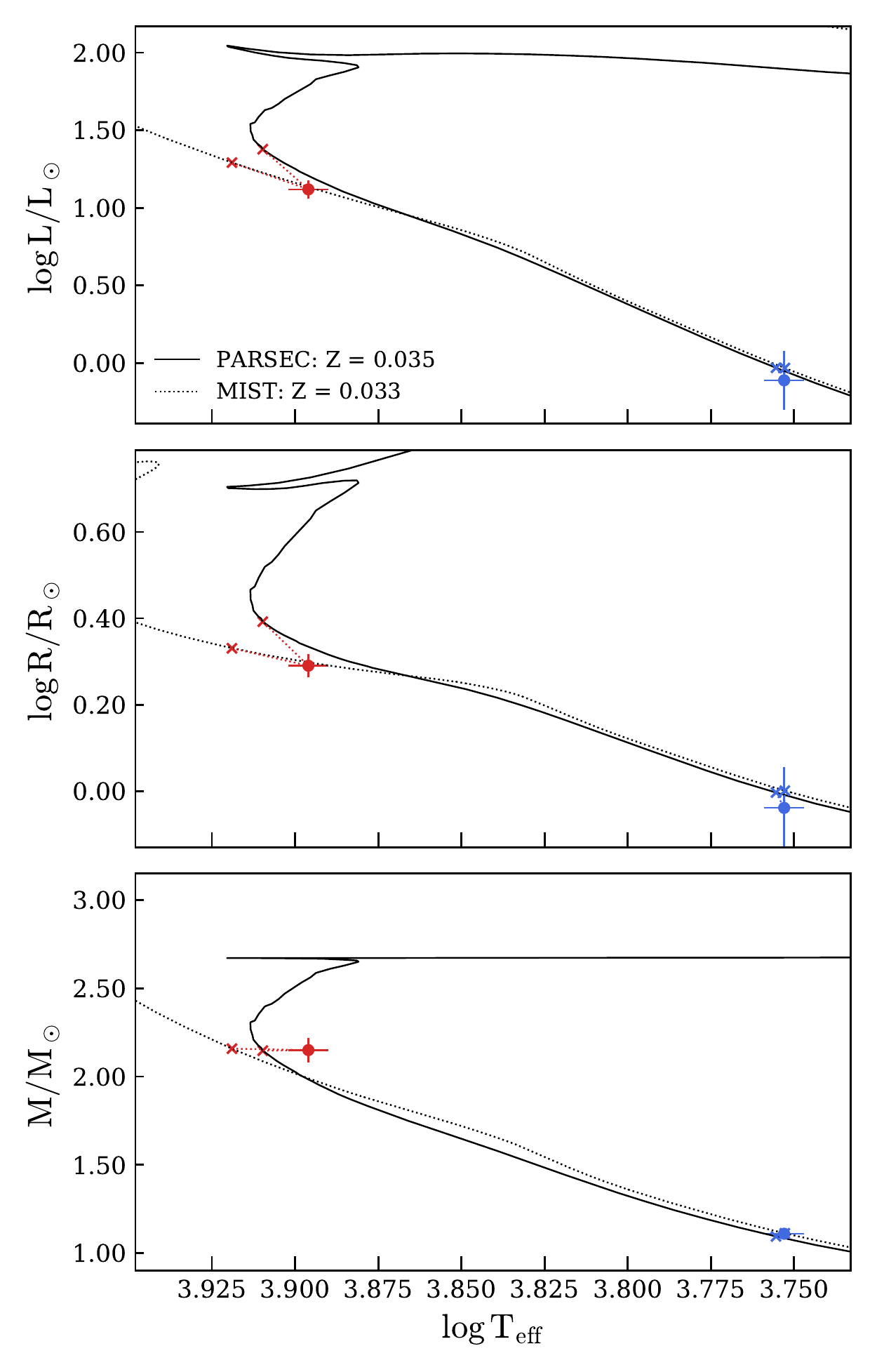}}
		\caption{Fitted \parsec and \mist isochrones for the V4090~Sgr system. We note that the luminosity is not fitted and is estimated from the Stefan-Boltzmann law.}
		\label{figure_isochrones_v4090sgr}
	\end{figure}

	We adopted the metallicity of $0.36\pm0.07$ estimated by \citet{North_1997_08_0} for the primary, as well as the effective temperature and radii. We note that their temperatures were estimated from average colour indices and not from spectroscopy. Here we only fitted the \parsec and \mist isochrones as we are out of the \basti metallicity range with this system. The isochrones displayed in Fig.~\ref{figure_isochrones_v4090sgr} show that both components are on the main sequence phase, with the primary star approaching the turn-off point. Both models provide a similar age (within $1\sigma$), although the \mist models give a younger system. We estimated $t_\mathrm{avg} = 0.44\pm0.13$\,Gyr. We notice that the observables of the secondary star are within $1\sigma$ with the isochrones, but this is not the case for the more-evolved component. The models predict a hotter component with a larger radii for this given mass, which might be linked to the core-overshooting or the calibration of isochrones. The disagreement could also be due to a bias in the Gaia parallax. To reconcile the isochrones to $\sim1\sigma$ we would need to correct the Gaia measurements by $+0.25$\,mas. We note that the Hipparcos parallax has a low accuracy of $\sim8$\,\%, providing a primary mass precise to $\sim30$\,\%, which does not help in constraining the models. Finally, the lack of radial velocities likely impacts a reliable estimates of the semi-amplitude parameters, and therefore on the masses.
	
	\begin{sidewaystable*}[!ht]
		\centering
		\caption{Stellar parameters used for the age determinations, together with our fitted and adopted age of the systems.}
		\begin{tabular}{ccccccc|ccccc} 
			\hline
			\hline
			System & Star	& $R$\,\tablefootmark{a} &  $T_\mathrm{eff}$ &  $\log L/L_\odot$	& [Fe/H] & Ref.	& $t_\mathrm{parsec}$ & $t_\mathrm{basti}$ & $t_\mathrm{mist}$ & $t_\mathrm{avg}$ & \\
			& 		& ($R_\odot$)		&  ($K$)   &  &  (dex)	&  & (Gyr) & (Gyr) & (Gyr) & (Gyr) &  \\
			\hline
			\multirow{2}{*}{AI~Phe} &A & $1.841\pm0.017$  & $6175\pm150$ & $0.646\pm0.043$ & \multirow{2}{*}{$-0.14\pm0.10$} & \multirow{2}{*}{1,2} & \multirow{2}{*}{$4.34\pm0.31$} & \multirow{2}{*}{$3.94\pm0.09$} & \multirow{2}{*}{$4.33\pm0.32$} & \multirow{2}{*}{$4.20\pm0.19$} \\
			& B & $2.907\pm0.013$ & $5140\pm120$ & $0.724\pm0.041$ &  & & &  \\
			\hline
			\multirow{2}{*}{AL~Dor} &A & $1.121\pm0.010$ & $6016\pm110$ & $0.159\pm0.033$ & \multirow{2}{*}{$0.10\pm0.10$\,\tablefootmark{b}} & \multirow{2}{*}{3} & \multirow{2}{*}{$3.03\pm0.61$} & \multirow{2}{*}{$2.65\pm1.11$} & \multirow{2}{*}{$3.19\pm0.16$} & \multirow{2}{*}{$2.97\pm0.23$} \\
			& B & $1.118\pm0.010$ & $5976\pm110$ & $0.145\pm0.033$ &  & & &  \\
			\hline
			\multirow{2}{*}{KW~Hya} &A & $2.124\pm0.015$ & $8000\pm200$ & $1.221\pm0.044$ & \multirow{2}{*}{$0.25\pm0.10$\,\tablefootmark{b}} & \multirow{2}{*}{3} & \multirow{2}{*}{$0.61\pm0.03$} & \multirow{2}{*}{$0.49\pm0.03$} & \multirow{2}{*}{$0.53\pm0.04$} & \multirow{2}{*}{$0.54\pm0.05$} \\
			& B & $1.439\pm0.022$ & $6960\pm210$ & $0.640\pm0.054$ &  & & &  \\
			\hline
			\multirow{2}{*}{NN~Del} &A & $2.194\pm0.015$ & $6437\pm200$ & $0.868\pm0.054$ & \multirow{2}{*}{$-0.10\pm0.10$\,\tablefootmark{b}} & \multirow{2}{*}{4} & \multirow{2}{*}{$2.40\pm0.17$} & \multirow{2}{*}{$1.96\pm0.22$} & \multirow{2}{*}{$2.28\pm0.15$} & \multirow{2}{*}{$2.21\pm0.18$} \\
			& B & $1.608\pm0.014$ & $6500\pm200$ & $0.615\pm0.054$ &  & & &  \\
			\hline
			\multirow{2}{*}{$\psi$~Cen} &A & $3.814\pm0.007$ & $10450\pm300$ & $2.147\pm0.050$ & \multirow{2}{*}{$0.05\pm0.10$\,\tablefootmark{b}} & \multirow{2}{*}{5,6} & \multirow{2}{*}{$0.29\pm0.02$} & \multirow{2}{*}{$0.27\pm0.02$} & \multirow{2}{*}{$0.28\pm0.01$} & \multirow{2}{*}{$0.28\pm0.01$} \\
			& B & $1.896\pm0.004$ & $8800\pm300$ & $1.244\pm0.059$ &  & & &  \\
			\hline
			\multirow{2}{*}{V4090~Sgr} &A & $1.95\pm0.13$ & $7700\pm100$ & $1.13\pm0.08$ & \multirow{2}{*}{$0.36\pm0.07$} & \multirow{2}{*}{7} & \multirow{2}{*}{$0.58\pm0.06$} & \multirow{2}{*}{--} & \multirow{2}{*}{$0.31\pm0.11$} & \multirow{2}{*}{$0.44\pm0.13$} \\
			& B & $0.91\pm0.21$ & $5800\pm80$ & $-0.02\pm0.21$ &  & & &  \\
			\hline
		\end{tabular}
		\tablefoot{References: 1- \citet{Graczyk_2017_03_0}. 2- \citet{Andersen_1988_05_0}. 3- \citet{Graczyk_2019_02_0}. 4- \citet{Sybilski_2018_07_0}. 5- \citet{Bruntt_2006_09_0}. 6- \citet{Mantegazza_2010_03_0}. 7- \citet{North_1997_08_0}.
			\tablefoottext{a}{Rescaled values according to our measured linear semi-major axis (see Sect.~\ref{section__discussion}).}
			\tablefoottext{b}{Adopted value (see text in Sect.~\ref{section__discussion}).}
		}
		\label{table__atmospheric_parameter}
	\end{sidewaystable*}

	In general, for all our systems we see that the models cannot properly reproduce all observables, especially the radii and masses. We do not see a general trend related to the metallicity, although we still lack precise measurements to reliably conclude. We however noticed that the stars located at the main-sequence stage are generally better fitted than the one located at the turn-off point. A possible explanation would be the parametrization of the convective core overshooting. For instance for KW~Hya and $\psi$~Cen which have stars at a similar evolution stage, we noticed that the model with the largest core overshooting parametrization (\parsec) gives a better match for both stars to all observables (assuming a metallicity of 0.05\,dex for $\psi$~Cen). Unfortunately, for a finer analysis, we will need metallicity measurements with a higher precision, and be able to enter the models to change specific internal stellar parameters to better compare between models instead of pre-computed isochrones.

	\subsection{Distance comparison with Gaia parallaxes}

\begin{figure}[!h]
		\centering
		\resizebox{\hsize}{!}{\includegraphics[width = \linewidth]{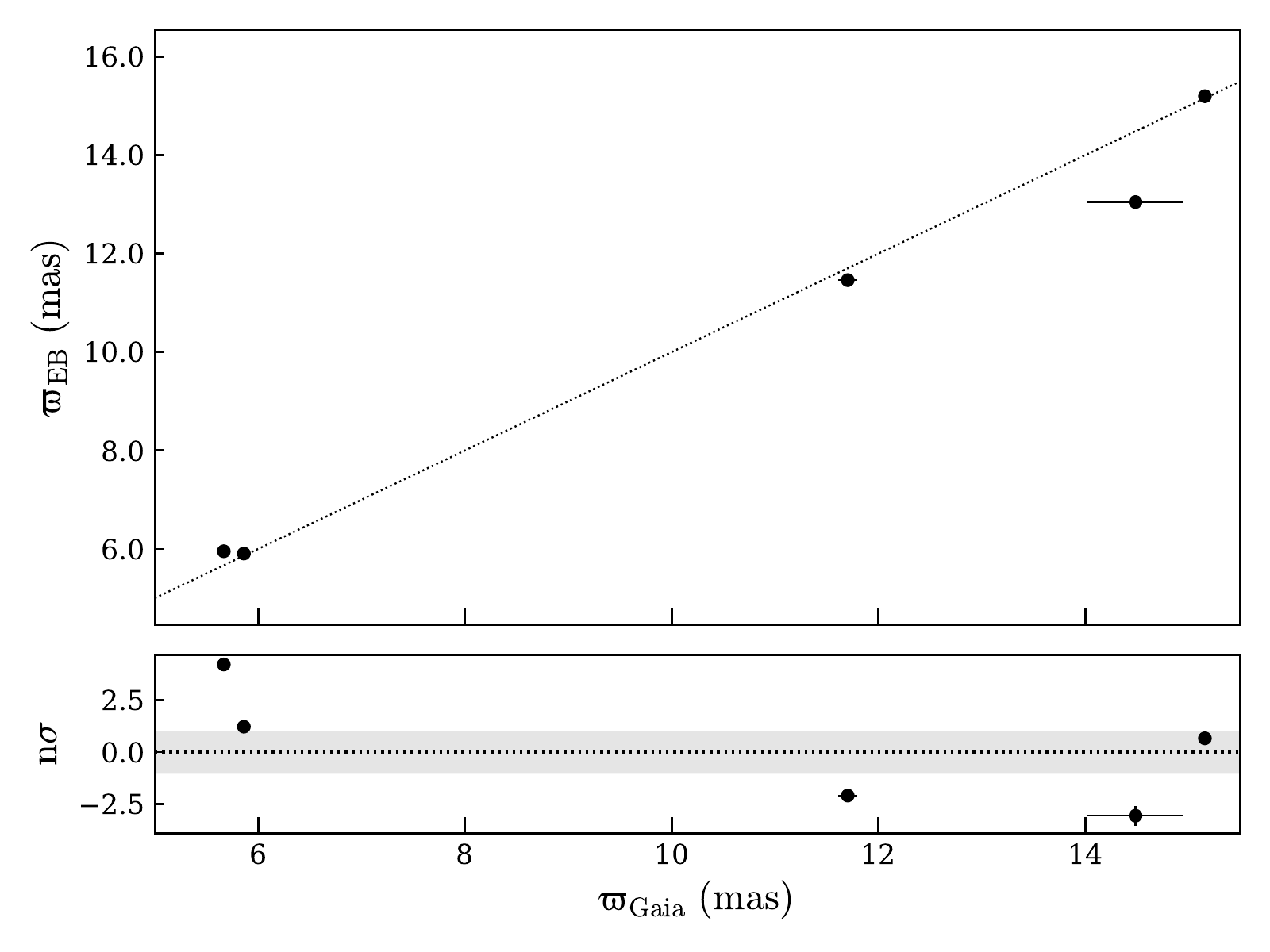}}
		\caption{Comparison between our derived orbital parallaxes and the Gaia second data release. The bottom panel shows the difference in number of $\sigma$, with the grey area representing $1\sigma$.}
		\label{figure_parallaxes}
	\end{figure}

	We displayed in Fig.~\ref{figure_parallaxes} the comparison between our derived orbital parallaxes and the ones from the GDR2. Only one system (AL~Dor) is in agreement within $1\sigma$. AI~Phe is at $1.2\sigma$ while the other systems are $>2\sigma$. We note that we took into account the $+29\,\mu$as astrometric offset suggested by \citet{Lindegren_2018_08_0}. As stated previously, this is to be expected as the GDR2 astrometric pipeline processing does not take into account yet the binarity of stars. However, it is worth mentioning that the Gaia bias indicators (the renormalized unit weight error and the astrometric excess noise) do not show strong evidence of possible biased measurements.

	We also examined the proper motion (PM) of these systems by comparing the Hipparcos and Gaia measurements following the approach developed by \citet[][see also \citealt{Kervella_2019_03_2}]{Kervella_2019_03_0}. The long time baseline of 24.25\,years between the two missions enables us to compare the mean long-term PM vectors with the individual measurements. Any difference (or anomaly) between the vectors indicates the presence of an orbiting third component. The detection is quantified using the signal-to-noise ratio (SNR) of the PM anomalies with respect to the mean PM as defined by \citet{Kervella_2019_03_2}. We found strong evidence of a third orbiting component for the AI~Phe and KW~Hya systems, with a SNR of 14.8 and 36.4 respectively. We note that this is consistent with our analysis of Sect.~\ref{section__results_for_individual_systems} for a third component of AI~Phe. A moderate evidence is found for V4090~Sgr and NN~Del (SNR $\sim 3$), while there is no sign of an additional component around $\psi$~Cen and AL~Dor.

	\section{Conclusion}
	\label{section__conclusion}
	
	This work reports the first interferometric observations of the eclipsing binary systems AI~Phe, AL~Dor, KW~Hya, NN~Del, $\psi$~Cen and V4090~Sgr using the VLTI/PIONIER combiner. We performed a simultaneous fit of our precise astrometric positions provided by interferometry with radial velocity measurements to measure all the orbital elements, accurate dynamical masses and orbital parallaxes. For some systems we obtained a precision down to 0.04\,\% on the masses and 0.4\,\% on the orbital parallaxes. The accuracy on the orbital parallaxes is dominated by the systematic uncertainty of 0.35\% on the interferometric separation measurements, limited by the knowledge of the effective wavelength of the PIONIER instrument. For our next targets of our observing program, such limitation will be overcome with the use of the VLTI/GRAVITY instrument \citep{Eisenhauer_2011_03_0}, which has a dedicated internal reference laser source providing a wavelength accuracy of  $\sim0.01$\,\%. Improving the precision on the astrometric and radial velocity measurements should also improved the precision on the masses.
	
	We confronted our precisely measured masses to stellar evolution models, together with stellar effective temperatures and radii retrieved from the literature. While our combined fit are satisfactory to derive the age of the systems, the high accuracy level on the masses of some systems showed that stellar evolution theory is still deficient in fitting simultaneously all parameters, with the radii or masses in marginal agreement with the models. This stresses the importance of precise stellar measurements for stellar evolution modelling and their calibration. Unfortunately, with precomputed isochrones (and predefined evolution parameters) and without precise measurements of the metallicity and temperature, it is impossible to conclude about which specific evolution model is more appropriate.
	
	With a median accuracy of $\sim 55\mu$as, our parallax measurements shows that the Gaia parallaxes still suffer from large systematic errors. However, this is expected as the last data release did not take into account the effect of binarity in the astrometric solutions. Such precise and accurate measurements provide a unique benchmark to test the next Gaia parallaxes.
	

	
	\begin{acknowledgements}
		The authors would like to thank all the people involved in the VLTI project. A.G. acknowledges support from FONDECYT grant 3130361. We acknowledge the support of the French Agence Nationale de la Recherche (ANR-15-CE31-0012-01, project Unlock-Cepheids). WG and GP gratefully acknowledge financial support from the BASAL Centro de Astrofisica y Tecnologias Afines (CATA, AFB-170002). WG also acknowledges financial support from the Millenium Institute of Astrophysics (MAS) of the Iniciativa Cientifica Milenio del Ministerio de Economia, Fomento y Turismo de Chile (project IC120009). We acknowledge financial support from the Programme National de Physique Stellaire (PNPS) of CNRS/INSU, France. Support from the Polish National Science Centre grants MAESTRO UMO-2017/26/A/ST9/00446 and from the IdP II 2015 0002 64 grant of the Polish Ministry of Science and Higher Education is also acknowledged. The research leading to these results has received funding from the European Research Council (ERC) under the European Union's Horizon 2020 research and innovation programme (grant agreement N$^\circ$695099, 639889 and 670519). BP acknowledges financial support from the Polish National Science Center grant SONATA 2014/15/D/ST9/02248. MT acknowledges financial support from the Polish National Science Center grant PRELUDIUM 2016/21/N/ST9/03310. We also thank Monica Reiner for kindly providing the reduced FEROS spectra of $\psi$~Cen. The research leading to these results has (partially) received funding from the KU\,Leuven Research Council (grant C16/18/005: PARADISE), from the Research Foundation Flanders (FWO) under grant agreement G0H5416N (ERC Runner Up Project), as well as from the BELgian federal Science Policy Office (BELSPO) through PRODEX grant PLATO.
	\end{acknowledgements}
	
	
	\bibliographystyle{aa}   
	\bibliography{/Users/agallenn/Sciences/Articles/bibliographie}
	
	
	\begin{appendix} 
		
		\section{Radial velocities}

		\begin{table*}[!ht]
	\centering
	\caption{Radial velocities of the AI~Phe system from the HARPS echelle spectrograph. Errorbars were rescaled according to the average scatter. Velocities listed here are corrected from the effect of the third component.}
	\begin{tabular}{ccccc|ccccc} 
		\hline
		\hline
		MJD\tablefootmark{a}		&	$V_1$	&	$\sigma_\mathrm{V_1}$	&  $V_2$	&	$\sigma_\mathrm{V_2}$& MJD\tablefootmark{a} &	$V_1$	&	$\sigma_\mathrm{V_1}$	&  $V_2$	&	$\sigma_\mathrm{V_2}$	\\
		(days)			&	(\kms)	&	(\kms)	& (\kms) & (\kms)	& (days) &	(\kms)	&	(\kms)	& (\kms) & (\kms)	 \\
		\hline
		55721.33272 & 26.057 & 0.032 & -29.181 & 0.026 &  56179.21858 & -53.641 & 0.032 & 47.335 & 0.026 \\
		55721.37130 & 26.333 & 0.032 & -29.456 & 0.026 &  56179.33669 & -53.073 & 0.032 & 46.785 & 0.026 \\
		55722.33576 & 32.991 & 0.032 & -35.807 & 0.026 &  56635.02474 & 45.041 & 0.032 & -47.427 & 0.026 \\
		55811.17302 & -48.736 & 0.032 & 42.667 & 0.026 &  56636.04064 & 45.538 & 0.032 & -47.894 & 0.026 \\
		55811.29172 & -47.931 & 0.032 & 41.894 & 0.026 &  56885.32747 & 20.233 & 0.032 & -23.559 & 0.026 \\
		55811.36526 & -47.408 & 0.032 & 41.399 & 0.026 &  56888.24509 & -31.994 & 0.032 & 26.596 & 0.026 \\
		55812.17541 & -41.070 & 0.032 & 35.269 & 0.026 &  57005.20377 & 45.143 & 0.032 & -47.509 & 0.026 \\
		55812.26846 & -40.283 & 0.032 & 34.515 & 0.026 &  57006.01155 & 42.507 & 0.032 & -44.952 & 0.026 \\
		55812.36954 & -39.415 & 0.032 & 33.688 & 0.026 &  57068.01308 & -26.313 & 0.032 & 21.125 & 0.026 \\
		55813.18472 & -32.202 & 0.032 & 26.770 & 0.026 &  57069.01599 & -16.708 & 0.032 & 11.983 & 0.026 \\
		55813.32021 & -30.947 & 0.032 & 25.576 & 0.026 &  57332.09583 & -48.074 & 0.032 & 42.058 & 0.026 \\
		56137.23379 & 8.524 & 0.032 & -12.219 & 0.026 &  57637.24896 & 7.616 & 0.032 & -11.396 & 0.026 \\
		56137.31727 & 9.258 & 0.032 & -12.932 & 0.026 &  57640.26189 & 31.244 & 0.032 & -34.154 & 0.026 \\
		56138.23715 & 16.980 & 0.032 & -20.457 & 0.026 &  57707.07373 & -29.479 & 0.032 & 24.140 & 0.026 \\
		56138.36237 & 18.005 & 0.032 & -21.452 & 0.026 &  57708.13134 & -19.423 & 0.032 & 14.508 & 0.026 \\
		56178.27474 & -56.476 & 0.032 & 50.075 & 0.026 &  58098.05945 & -50.001 & 0.032 & 43.768 & 0.026 \\
		56179.12360 & -54.078 & 0.032 & 47.760 & 0.026 &  & & & & \\
		\hline
	\end{tabular}
	\tablefoot{\tablefoottext{a}{HJD-2~400~000.5}}
	\label{table_rv_aiphe}
	\end{table*}

		\begin{table*}[!ht]
			\centering
			\caption{Radial velocities of the NN~Del system from the STELLA echelle spectrograph.}
			\begin{tabular}{ccccc|ccccc} 
				\hline
				\hline
				MJD\tablefootmark{a}		&	$V_1$	&	$\sigma_\mathrm{V_1}$	&  $V_2$	&	$\sigma_\mathrm{V_2}$& MJD\tablefootmark{a} &	$V_1$	&	$\sigma_\mathrm{V_1}$	&  $V_2$	&	$\sigma_\mathrm{V_2}$	\\
				(days)			&	(\kms)	&	(\kms)	& (\kms) & (\kms)	& (days) &	(\kms)	&	(\kms)	& (\kms) & (\kms)	 \\
				\hline
				56733.21347 & -26.202 & 0.100 & 8.495 & 0.180 &  56891.91890 & 27.283 & 0.100 & -49.451 & 0.180 \\
				56734.23318 & -26.266 & 0.100 & 8.683 & 0.180 &  57452.22954 & -24.679 & 0.100 & 7.198 & 0.180 \\
				56735.21681 & -26.493 & 0.100 & 8.778 & 0.180 &  57457.25205 & -22.298 & 0.100 & 4.554 & 0.180 \\
				56740.20810 & -27.125 & 0.100 & 9.551 & 0.180 &  57459.25730 & -20.836 & 0.100 & 2.948 & 0.180 \\
				56748.16840 & -27.058 & 0.100 & 9.229 & 0.180 &  57474.24531 & 11.560 & 0.100 & -32.166 & 0.180 \\
				56753.15930 & -26.362 & 0.100 & 8.548 & 0.180 &  57475.24264 & 16.394 & 0.100 & -37.472 & 0.180 \\
				56776.15487 & -1.122 & 0.100 & -19.005 & 0.180 &  57476.24249 & 21.724 & 0.100 & -43.196 & 0.180 \\
				56811.99682 & -15.729 & 0.100 & -2.475 & 0.180 &  57477.23993 & 27.291 & 0.100 & -49.322 & 0.180 \\
				56815.99358 & -18.902 & 0.100 & 0.957 & 0.180 &  57479.19131 & 37.629 & 0.100 & -60.917 & 0.180 \\
				56819.98256 & -21.404 & 0.100 & 3.605 & 0.180 &  57482.23986 & 45.242 & 0.100 & -68.988 & 0.180 \\
				56823.97850 & -23.230 & 0.100 & 5.614 & 0.180 &  57491.14149 & 12.980 & 0.100 & -34.008 & 0.180 \\
				56829.12482 & -25.030 & 0.100 & 7.664 & 0.180 &  57495.16072 & 1.505 & 0.100 & -21.003 & 0.180 \\
				56842.94818 & -27.083 & 0.100 & 9.928 & 0.180 &  58071.89626 & 21.787 & 0.100 & -43.601 & 0.180 \\
				56859.04479 & -23.823 & 0.100 & 6.183 & 0.180 &  58075.88572 & 42.141 & 0.100 & -65.805 & 0.180 \\
				56861.98510 & -22.114 & 0.100 & 4.358 & 0.180 &  58077.89531 & 45.066 & 0.100 & -69.110 & 0.180 \\
				56863.97888 & -20.568 & 0.100 & 2.674 & 0.180 &  58167.27136 & 3.479 & 0.100 & -23.386 & 0.180 \\
				56866.01320 & -18.660 & 0.100 & 0.644 & 0.180 &  58168.26857 & 7.587 & 0.100 & -27.912 & 0.180 \\
				56868.03402 & -16.272 & 0.100 & -1.957 & 0.180 &  58171.26807 & 22.439 & 0.100 & -44.761 & 0.180 \\
				56879.96034 & 18.000 & 0.100 & -39.396 & 0.180 &  58372.05146 & 34.908 & 0.100 & -57.754 & 0.180 \\
				56886.14830 & 45.068 & 0.100 & -69.053 & 0.180 &  58380.08501 & 31.112 & 0.100 & -53.504 & 0.180 \\
				\hline
			\end{tabular}
		\tablefoot{\tablefoottext{a}{HJD-2~400~000.5}}
			\label{table_rv_nndel}
		\end{table*}
		
		\begin{table*}[!ht]
			\centering
			\caption{Radial velocities of the $\psi$~Cen system.}
			\begin{tabular}{cccccc} 
				\hline
				\hline
				MJD\tablefootmark{a}		&	$V_1$	&	$\sigma_\mathrm{V_1}$	&  $V_2$	&	$\sigma_\mathrm{V_2}$ & Inst.	\\
				(days)			&	(\kms)	&	(\kms)	& (\kms) & (\kms)	&	 \\
				\hline
                53833.04211  &  13.923  &  0.640 &   --       &  --  &  FEROS \\
				53834.15257  &  20.228  &  0.640 &   -23.043  &  1.570  &  FEROS \\
				53835.20515  &  27.815  &  0.640 &   --        &  --  &  FEROS \\
				53838.23353  &  64.721  &  0.640 &   -96.028  &  1.570 &  FEROS \\
				53868.16063  &  -2.649  &  0.640 &   12.848    & 1.570  &  FEROS \\
				53883.24819  &  0.915   &  0.640 &   7.582    &  1.570 &  FEROS \\
				53883.25006  &  1.903   &  0.640 &   6.144    &  1.570 &  FEROS \\
				53892.97575  &  -21.112  & 0.640  &   43.259  &  1.570 &  FEROS \\
				53898.11363  &  -17.142  & 0.640  &   37.151  &  1.570 &  FEROS \\
				53900.13525  &  -16.239  & 0.640  &   34.071  &  1.570 &  FEROS \\
				53904.07395  &  -10.736  & 0.640  &   25.864  &  1.570 &  FEROS \\
				53918.02272  &  72.353  &  0.640 &   -108.351  & 1.570  &  FEROS \\
				53918.99671  &  51.314  &  0.640 &   -74.643  &  1.570 &  FEROS \\
				53954.96673  &  68.935  &  0.640 &   -102.158  & 1.570  &  FEROS \\
				53956.14504  &  77.999  &  0.640 &   -117.662  & 1.570  &  FEROS \\
				53956.97084  &  70.110  &  0.640 &   -104.681  & 1.570  &  FEROS \\
				53963.97849  &  -13.988  & 0.640  &   31.493  &  1.570 &  FEROS \\
				53985.99695  &  2.972   &  0.640 &   4.775     & 1.570  &  FEROS \\
				53991.96071  &  43.798  &  0.640 &   --        &  --  &  FEROS \\
				53991.96542  &  45.454  &  0.640 &   -60.994  &  1.570  &  FEROS \\
				53991.97679  &  43.689  &  0.640 &   -66.196  &  1.570  &  FEROS \\
				53995.97828  &  66.342  &  0.640 &   -98.152  &  1.570  &  FEROS \\
				53997.98533  &  22.587  &  0.640 &   -24.478  &  1.570  &  FEROS \\
				53997.98897  &  23.025  &  0.640 &   -29.820  &  1.570  &  FEROS \\
				53997.99360  &  22.829  &  0.640 &   -27.586  &  1.570  &  FEROS \\
				53997.99770  &  22.205  &  0.640 &   -28.004  &  1.570  &  FEROS \\
				53998.00185  &  22.955  &  0.640 &   -27.446  &  1.570  &  FEROS \\
				53998.00686  &  22.474  &  0.640 &   -31.516  &  1.570  &  FEROS \\
				53998.01098  &  23.210  &  0.640 &   -24.092  &  1.570  &  FEROS \\
				53998.01511  &  22.850  &  0.640 &   -27.787  &  1.570  &  FEROS \\
				53998.02256  &  22.606  &  0.640 &   -24.960  &  1.570  &  FEROS \\
				53998.02668  &  21.618  &  0.640 &   -27.619  &  1.570  &  FEROS \\
				53998.03080  &  22.085  &  0.640 &   -26.103  &  1.570  &  FEROS \\
				53998.99093  &  10.075  &  0.640 &   -4.221   &  1.570  &  FEROS \\
				54005.98158  &  -19.883  & 0.640  &   38.664  &  1.570  &  FEROS \\
				56876.96871  &  -17.302  & 0.640  &   37.173  &  1.570  &  HARPS \\
				56877.96639  &  -18.613  & 0.640  &   39.958  &  1.570  &  HARPS \\
				56878.96818  &  -19.415  & 0.640  &   42.070  &  1.570  &  HARPS \\
				56907.97716  &  40.338  &  0.640 &   -60.257  &  1.570  &  HARPS \\
				56908.97554  &  21.467  &  0.640 &   -23.608  &  1.570  &  HARPS \\
				56909.98788  &  8.432   &  0.640 &   -3.744   &  1.570  &  HARPS \\
				57029.35971  &  -11.754  & 0.640  &   29.700  &  1.570  &  HARPS \\
				57030.37638  &  -15.066  & 0.640  &   32.946  &  1.570  &  HARPS \\
				\hline
			\end{tabular}
		\tablefoot{\tablefoottext{a}{HJD-2~400~000.5}}
			\label{table_rv_psicen}
		\end{table*}
		
		\section{Parameters of the calibrators used for the PIONIER interferometric observations}
		\label{appendix__calibrators}
		
		\begin{table*}[!ht]
			\centering
			\caption{Calibrators used for our observations for all systems.}
			\begin{tabular}{cccccc|cccccc} 
				\hline
				\hline
				\# & HD		&	Sp.~type	&	$V$  & $H$ 	&  $\theta_\mathrm{UD}$  & \#  &  HD		&	Sp.~type	&	$V$  & $H$ 	&  $\theta_\mathrm{UD}$	\\
				&	&					&	(mag)	&	(mag)	&	(mas)				&			&	&	&	(mag)	&	(mag)	&	(mas)   \\
				\hline
				\multicolumn{6}{c|}{\object{AI~Phe}} &  \multicolumn{6}{c}{\object{AL~Dor}}\\
				1	& 	\object{HD5311}  &  F7V  &  7.91  & 6.85  & $0.178\pm0.004$  &	1	&  \object{HD27917}  &  F3IV  &  7.91  & 6.90  & $0.172\pm0.012$  \\
				2	&	\object{HD9404}  &  F3V  &  7.86  & 6.90  & $0.169\pm0.004$  &	2	&  \object{HD28421}  &  G1V  &  8.24  & 6.98  & $0.175\pm0.005$  \\
				3	&	\object{HD4025}  &  F7V  &  8.14  & 6.89  & $0.176\pm0.004$  &	3	&  \object{HD30880}  &  F3IV  &  7.87  & 6.91  & $0.169\pm0.004$  \\
				4	&	\object{HD7211}  &  K0III  &  9.11  & 6.84  & $0.217\pm0.005$  &	4	&  \object{HD32363}  &  F7/8V  &  8.14  & 6.87  & $0.179\pm0.004$  \\
				5	&	\object{HD5386}  &  K0III &  8.92  & 6.85  & $0.209\pm0.006$  &	   & & & & &  \\
				6	&	\object{HD8640}  &  K0III  &  9.22  & 6.93  & $0.208\pm0.005$  &   & & & & &  \\
				\hline
				\multicolumn{6}{c|}{\object{NN~Del}} &  \multicolumn{6}{c}{\object{KW~Hya}}\\
				1	& 	\object{HD197449}  & G8III/IV   &  8.60  & 6.52  & $0.242\pm0.006$  &	1	&  \object{HD81342}  &  F3III/IV  &  6.92  & 5.98  & $0.267\pm0.019$  \\
				2	&	\object{HD196201}  &  G5  &  8.48  & 6.70  & $0.215\pm0.005$  &	2	&  \object{HD82043}  &  F0III  &  6.13  & 5.59  & $0.275\pm0.019$  \\
				3	&	\object{HD197195}  &  G5  &  8.24  & 6.95  & $0.183\pm0.004$  &	3	&  \object{HD83712}  &  F8IV/V  &  7.04  & 5.98  & $0.272\pm0.019$  \\
				\hline
				\multicolumn{6}{c|}{\object{$\Psi$~Cen}} &  \multicolumn{6}{c}{\object{V4090~Sgr}}\\
				1	& 	\object{HD122641}  & K0III   &  7.90  & 5.54  & $0.408\pm0.029$  &	  1	&  \object{HD188049}  &  G6V  &  8.50  & 6.56  & $0.234\pm0.017$  \\
				2	&	\object{HD129217}  & K0III  &  7.53  & 5.35  & $0.423\pm0.030$  & 2 & \object{HD187369}  &  G2IV  &  7.86  & 6.53  & $0.217\pm0.015$  \\
				3	&	\object{HD128555}  &  K0III  & 7.60  & 5.44  & $0.396\pm0.028$  & 3 &	 \object{HD186085}  &  G3V  &  8.40  & 6.88  & $0.199\pm0.014$  \\
				
				\hline
			\end{tabular}
			\label{table_calibrators}
		\end{table*}
		
	\end{appendix}
	
\end{document}